\documentclass[%
  aps,
  prd,
  reprint,
  amsmath,
  amssymb,
  superscriptaddress, 
  nofootinbib]{revtex4-2}

\usepackage[utf8]{inputenc}
\usepackage[english]{babel}

\usepackage{mathptmx}
\usepackage{graphicx}\graphicspath{{images/}}
\usepackage{upgreek}
\usepackage{xcolor}
\definecolor{darkblue}{rgb}{0,0,0.5}
\definecolor{firebrick}{rgb}{0.75,0.125,0.125}
\definecolor{darkgreen}{rgb}{0,0.5,0}
\usepackage[colorlinks=true,linkcolor=firebrick,citecolor=darkgreen,urlcolor=darkblue]{hyperref}
\usepackage[capitalise,nameinlink]{cleveref}

\usepackage[separate-uncertainty, multi-part-units=single]{siunitx}
\usepackage[caption=false]{subfig}

\usepackage{xspace}
\usepackage{tabularx}
\usepackage{booktabs}
\usepackage{adjustbox}
\DeclareRobustCommand*{\rmuM}{\ensuremath{F_\upmu}}
\DeclareRobustCommand*{\xmaxM}{\ensuremath{X_\text{max}}}
\DeclareRobustCommand*{\fisherM}{\ensuremath{f}}

\newcommand{\rmuT}{$\rmuM$\xspace}

\newcommand{\xmaxT}{$\xmaxM$\xspace}

\newcommand{\fisherT}{$\fisherM$\xspace}

\newcommand{\tabEffNam}{$\varepsilon$ (\%)}
\newcommand{\calE}{\mathcal{E}}

\DeclareSIUnit\gauss{G}
\DeclareSIUnit\VEM{VEM}
\DeclareSIUnit\parsec{pc}
\DeclareSIUnit\year{yr}
\begin{document}
\preprint{PRD/123-QED}

\title{Search for photons above \texorpdfstring{10$^{18}$}{10\^18}  eV by simultaneously measuring the atmospheric depth and the muon content of air showers at the Pierre Auger Observatory}

\author{A.~Abdul Halim}
\affiliation{University of Adelaide, Adelaide, S.A., Australia}

\author{P.~Abreu}
\affiliation{Laborat\'orio de Instrumenta\c{c}\~ao e F\'\i{}sica Experimental de Part\'\i{}culas -- LIP and Instituto Superior T\'ecnico -- IST, Universidade de Lisboa -- UL, Lisboa, Portugal}

\author{M.~Aglietta}
\affiliation{Osservatorio Astrofisico di Torino (INAF), Torino, Italy}
\affiliation{INFN, Sezione di Torino, Torino, Italy}

\author{I.~Allekotte}
\affiliation{Centro At\'omico Bariloche and Instituto Balseiro (CNEA-UNCuyo-CONICET), San Carlos de Bariloche, Argentina}

\author{K.~Almeida Cheminant}
\affiliation{Nationaal Instituut voor Kernfysica en Hoge Energie Fysica (NIKHEF), Science Park, Amsterdam, The Netherlands}
\affiliation{IMAPP, Radboud University Nijmegen, Nijmegen, The Netherlands}
\affiliation{Institute of Nuclear Physics PAN, Krakow, Poland}

\author{A.~Almela}
\affiliation{Instituto de Tecnolog\'\i{}as en Detecci\'on y Astropart\'\i{}culas (CNEA, CONICET, UNSAM), Buenos Aires, Argentina}
\affiliation{Universidad Tecnol\'ogica Nacional -- Facultad Regional Buenos Aires, Buenos Aires, Argentina}

\author{R.~Aloisio}
\affiliation{Gran Sasso Science Institute, L'Aquila, Italy}
\affiliation{INFN Laboratori Nazionali del Gran Sasso, Assergi (L'Aquila), Italy}

\author{J.~Alvarez-Mu\~niz}
\affiliation{Instituto Galego de F\'\i{}sica de Altas Enerx\'\i{}as (IGFAE), Universidade de Santiago de Compostela, Santiago de Compostela, Spain}

\author{J.~Ammerman Yebra}
\affiliation{Instituto Galego de F\'\i{}sica de Altas Enerx\'\i{}as (IGFAE), Universidade de Santiago de Compostela, Santiago de Compostela, Spain}

\author{G.A.~Anastasi}
\affiliation{Universit\`a di Catania, Dipartimento di Fisica e Astronomia ``Ettore Majorana``, Catania, Italy}
\affiliation{INFN, Sezione di Catania, Catania, Italy}

% ...

\collaboration{The Pierre Auger Collaboration}
\email{spokespersons@auger.org}
\homepage{http://www.auger.org}
\noaffiliation


\date{\today}

\begin{abstract}
The Pierre Auger Observatory is the most sensitive instrument to detect photons with energies above $10^{17}$\,eV.
It measures extensive air showers generated by ultra high energy cosmic rays using a hybrid technique 
that exploits the combination of a fluorescence detector with a ground array of particle detectors.
The signatures of a photon-induced air shower are a larger atmospheric depth of the shower maximum
($X_\text{max}$) and a steeper lateral distribution function, along with a lower number of muons with respect to the bulk of hadron-induced cascades.
In this work, a new analysis technique in the energy interval between $1$ and \SI{30}{\exa\eV} (1\,EeV = $10^{18}$\,eV) has been developed
by combining the fluorescence detector-based measurement of $X_\text{max}$ with the specific features of the surface detector signal through a parameter related to the air shower muon content, derived from the universality of the air shower development.
No evidence of a statistically significant signal due to photon primaries was found 
using data collected in about 12 years of operation. Thus, upper bounds to the integral photon flux have been set using a detailed calculation of the detector exposure, in combination with a data-driven background estimation.
The derived 95\% confidence level upper limits 
are 0.0403, 0.01113, 0.0035, 0.0023, and 0.0021 km$^{-2}$ sr$^{-1}$ yr$^{-1}$ above 1, 2, 3, 5, and 10 EeV, respectively,  
leading to the most stringent upper limits on the photon flux in the EeV range. Compared with  past results, the upper limits were improved by about 40\% for the lowest energy threshold and by a factor 3 above 3 EeV, where no candidates were found and the expected background is negligible.
The presented limits can be used to probe
the assumptions on chemical composition of ultra-high energy cosmic rays and allow for the constraint of the mass and lifetime phase space of super-heavy dark matter particles.




\end{abstract}

\maketitle

\newpage

\section{Introduction}
\label{sec::intro}

Photons with energy higher than 1\,EeV = $10^{18}$\,eV are expected to be produced by ultra-high energy 
cosmic rays (UHECRs) in interactions with the microwave background radiation during 
their propagation to Earth, via the Greisen-Zatsepin-Kuzmin effect~\cite{Greisen,ZatsepinKuzmin}. 
The produced 
photons may, in turn, interact with the soft photons of the extra-galactic background light (EBL), resulting in a flux significantly lower than that of UHECRs by orders of magnitude and limiting the explored horizon to a few Mpc~\cite{PropUHECR}. 
Nevertheless implications of the search for EeV photons remain relevant for both cosmic-ray and fundamental physics. 
Unlike charged cosmic rays, which are deflected by the magnetic fields permeating the Universe, photons point back to their sources.
Therefore, the quest for the origin of UHECRs benefits from a multi-messenger approach, since direct information about their acceleration sites can be obtained 
by searching for the neutral particles, photons and neutrinos, 
generated by the interactions of cosmic rays at the acceleration sites, via the so-called \emph{astrophysical beam dump} process~\cite{gaisser1990, HalzenNatPh}. 
Cosmogenic photons can also probe UHECRs as their flux depends on the 
characteristics of the sources, as well as on the nature of the parent
nuclei. Finally, EeV photons might probe new physics, as their detection 
would be a smoking gun for dark matter composed of super-heavy particles 
decaying to photons or other exotic scenarios~\cite{ANCHORDOQUI2021102614,PhysRevD.107.042002,PhysRevLett.130.061001}.

Due to the weakness of both the cosmic-ray and cosmic-photon fluxes,
the photon search can presently only be done through large ground-based detectors that exploit the phenomenon of extensive air showers.
The identification of photon primaries
relies on the ability to
distinguish the showers generated by photons from those initiated by the overwhelming background of
protons and heavier nuclei. 
Since the radiation length in the atmosphere is more than two orders of magnitude smaller than the mean
free path for photo-nuclear interaction in the ultra-high energy range, in photon showers the transfer of energy to
the hadron/muon channel is much smaller than in hadron-induced air-showers,
resulting in a lower number
of secondary muons.
Additionally, as the development of photon showers is delayed by the
typically small multiplicity of electromagnetic interactions and even further in the EeV energy range due to the LPM effect~\cite{Landau:1953,*Landau:1953gr,*Migdal}, the depth of the shower maximum, $X_\text{max}$, is deeper in the atmosphere with respect to showers initiated by hadrons.

In this work, a search for photons at energies above 1\,EeV using the Auger data is presented.
The paper is structured as follows: \cref{sec::PAO} provides a brief description of the Pierre Auger Observatory and of its hybrid operating mode, combining a Surface Detector array (SD) with a Fluorescence Detector (FD). 
In \cref{sec::approach}, we will introduce a new method for calculating a parameter, \rmuT, related to the muonic component of an air shower. 
The method is based on air shower universality in combination with the high-quality reconstruction of hybrid events, simultaneously observed by the FD and the SD\@. 
The analysis is applied to 12 years of high-quality selected data, as discussed in \cref{sec::dataset}\@.
To fully exploit the hybrid approach, 
\rmuT is combined with the depth of the shower maximum, \xmaxT, measured by the Fluorescence Detector of the Pierre Auger Observatory in a Fisher discriminant analysis in \cref{sec::combination}\@. 
The expected background resulting from hadron-like showers is examined in \cref{sec::background}\@. 
The result of the application of the selection strategy to data is detailed in \cref{sec::results}\@. 
In the absence of any significant signal, we establish upper limits on the integral photon flux, which are presented in \cref{sec::discussion}, along with their associated systematic uncertainties. Finally, the implications of the derived results 
are briefly discussed in~\cref{sec::conclusions}\@.

\section{The Pierre Auger Observatory}
\label{sec::PAO}
The Pierre Auger Observatory~\cite{PierreAuger:2015eyc}, located near the town of Malarg\"ue in the Argentinian \emph{Pampa Amarilla}, 
is the largest cosmic-ray observatory to date, offering an unprecedented exposure to EeV photons. A key feature of the Pierre Auger Observatory is its hybrid concept, based on the combination of measurements provided by a surface detector array and a fluorescence detector. 
 The surface detector consists of 1600 water-Cherenkov detector (WCD) stations arranged on a triangular grid with a spacing of $\SI{1500}{\m}$,
covering a total area of about $\SI{3000}{\square\km}$. 

The SD samples the shower density at ground level, i.e., the distribution of particles as a function of the distance from the shower axis, with a duty cycle of 100\%. Moreover the time profiles of the signals recorded with the WCDs can be used to build
observables which are sensitive to the nature of the primary cosmic ray. The SD is overlooked by 27 fluorescence telescopes, located at four sites at the border of the array, with field of view of $30^{\circ} \times 30^{\circ}$. The FD records the longitudinal shower development in the atmosphere above the SD and it can only be operated during clear, moonless nights, reducing the duty cycle to 15\%. 
The FD provides a direct observation of the longitudinal shower profile, which allows for the measurement of the energy, $E$, and of the \xmaxT of a shower. Each fluorescence telescope hosts a camera of 440 photomultipliers (pixels). The pattern of triggered pixels in the telescopes along with their trigger times are used to reconstruct the geometry of the incoming showers. At this level, the temporal information provided by even a single station of the SD can greatly improve the accuracy in determining the shower direction and its impact point at the ground (hereafter named as core). Once the geometry is reconstructed, the energy deposited in the atmosphere by secondary particles can be derived through the measurement of the fluorescence light emitted by nitrogen molecules during the passage of the shower through the atmosphere. This is done by making use of the optical properties of the atmosphere provided by several instruments continuously monitoring the volume over the array, as described in~\cite{PierreAuger:2012,icrc2019_violet}. The energy of the primary particle is finally derived in a calorimetric way as the integral of the fit of a modified Gaisser-Hillas function to the observed longitudinal profile~\cite{UNGER2008433,usp2019}, and corrected for the invisible energy fraction carried by neutrinos and muons by following a data driven approach~\cite{PhysRevD.100.082003}. 


\section{Search for photons in the context of air shower Universality}
\label{sec::approach}
\begin{figure}[htbp]
\centering
\includegraphics[width=\columnwidth]{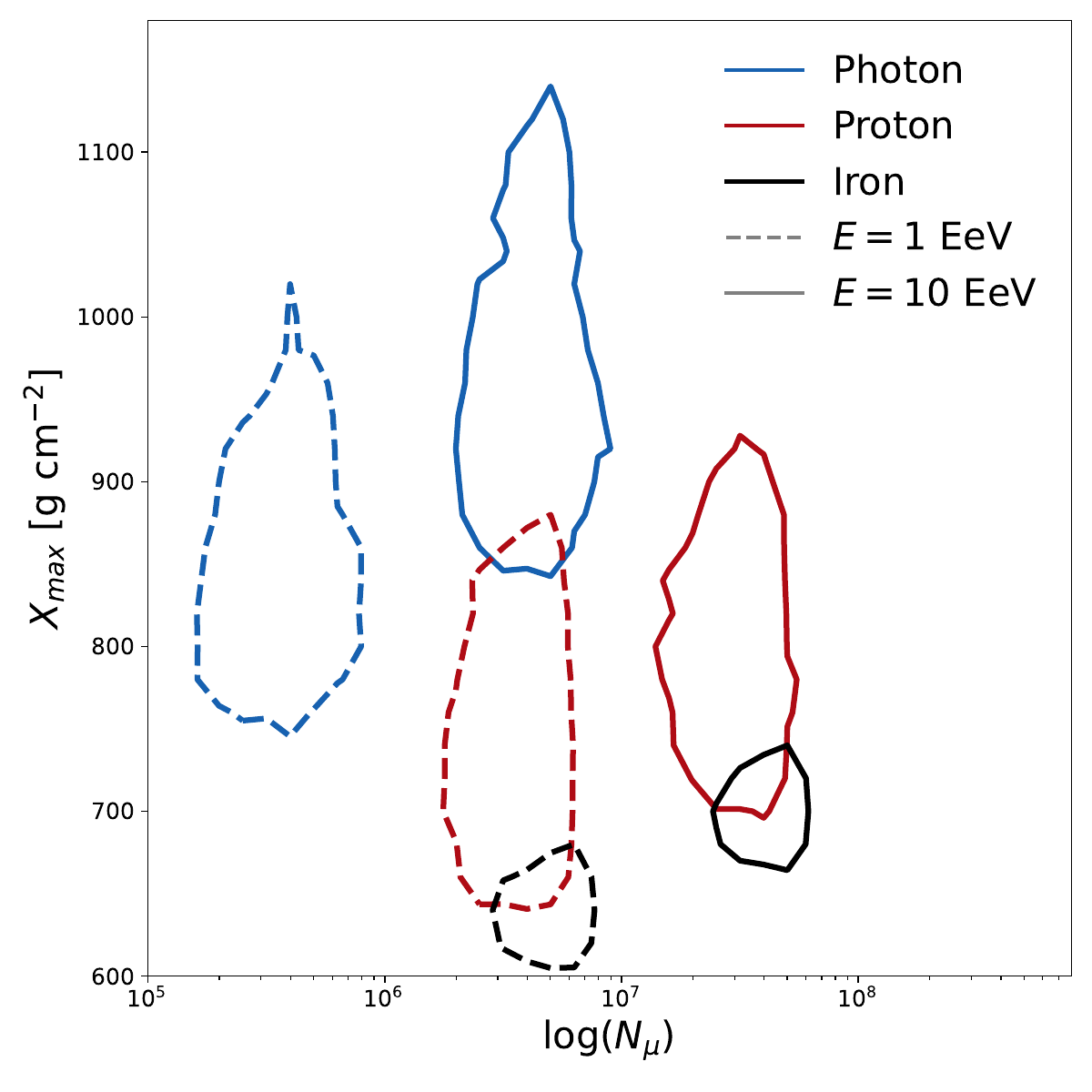}
\caption{
        \xmaxT and number of muons distributions for simulated air shower initiated by proton- (red) photon- (blue) iron- (black) primaries, at
        1\,EeV (dashed) and 10\,EeV (solid). 
        Contour lines enclose the 90\% of the distribution for each primary type. The CONEX~\cite{Conex2007} air-shower generator was used for this plot.}
\label{fig:PrimariesSeparation}
\end{figure}

In this work, we will perform the photon-identification using hybrid events, i.e., events detected by the FD in combination with the SD.
The main signature of a photon-induced air shower is a larger atmospheric depth at the shower maximum and a lower number of muons than the bulk of hadron-induced background. In \cref{fig:PrimariesSeparation}, the distributions of \xmaxT and the number of muons are shown for simulated air showers initiated by proton (red), photon (blue) and iron (black) primaries at 
1\,EeV and 10\,EeV.
Photon-initiated showers are well separated from showers initiated by hadrons in both cases. 
CONEX~\cite{Conex2007} air-shower generator was used for this plot.

\begin{figure}[!htb]
\centering
\includegraphics[width=\columnwidth]{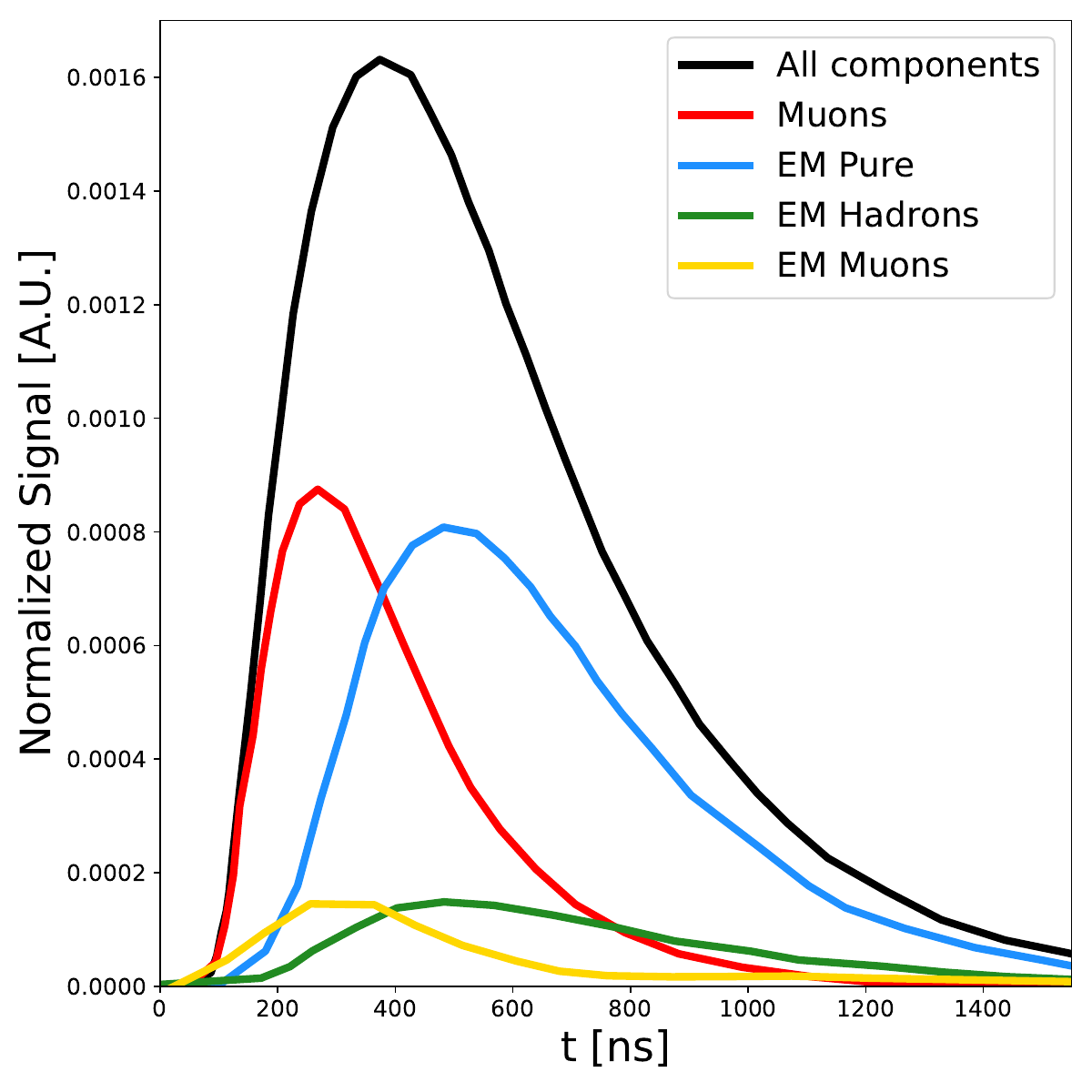}
\caption{
        Conceptual visualization of the surface detector signal components as a function of time, according to the universality paradigm~\cite{AVE201746}.}
        
    \label{fig:UniversalityTimeSignal}
\end{figure}

Although the SD observes showers at a fixed depth, the specific characteristics of the longitudinal development and the relative weight of shower particle components 
are embedded in the detected signals. 
To extract this information, we developed a new variable, \rmuT, related to the muonic content of the shower, derived in the context of air shower universality.  
The general idea behind air shower universality is that the energy spectrum of the secondary particles produced during the shower development, as well as their angular and lateral distributions, depend only on the energy of the primary and on the stage of the shower development~\cite{Lafebre:2009,Cazon:2023}. Thus the average properties of an EAS can be described by a few macroscopic shower characteristics. In general, for a given detector, it is possible to predict the signal produced by secondary particles at the different stages of the shower development. For the specific case of the SD detector, the total signal in each WCD can be modelled as the superposition of four components:  
$S_\upmu$ for muons, $S_{\text{e}\upgamma}$ for $\text{e}^\pm$ and $\upgamma$ from high-energy pions, $S_{\text{e}\upgamma(\upmu)}$ for $\text{e}^\pm$ and $\upgamma$
from muon decays, $S_{\text{e}\upgamma(\text{had})}$ for $\text{e}^\pm$ and $\upgamma$ due to low-energy
hadrons.
A parameterisation of each signal component for the WCD of the Pierre Auger Observatory 
was derived in~\cite{AVE201723, AVE201746}.
The relative contributions of the described four components to the expected overall signal of a surface detector station are visualized in~\cref{fig:UniversalityTimeSignal} for an exemplary simulated proton of about 30 EeV. 
The predicted total signal, $S_\text{pred}$, can be expressed as
\begin{equation}
  \label{eq:univ}
  S_\text{pred} =  \sum_{i=1}^4 S_i = \sum_{i=1}^4 f_i(\rmuM) \, S_{i,\text{comp}}
\end{equation}
where $i$ runs over the four components, 
$S_{i,\text{comp}}$ is the signal of each component that, according to the universality model, 
depends only on the primary total energy $E$, on \xmaxT, and on the geometry of the shower. 
The dependence on the mass of the primary particle is factorized and entirely contained in the terms denoted as $f_i$.

More specifically, the variable $\rmuM$ is defined as the ratio of the signal due to the muonic contribution $S_\upmu$, and its reference value $S_{\upmu,\text{ref}}$ calculated for a proton-induced shower in a way that a value of $\rmuM$ significantly less than one is indicative of a large deviation from a hadronic shower.  

The proposed method aims at maximizing the benefit of combining the predictive power of the universality model with the accuracy in the determination of $E$, \xmaxT and shower geometry provided by the hybrid reconstruction.  
Namely, $S_{i,\text{comp}}$ can be directly calculated for each station involved in a hybrid event using as input the hybrid-reconstructed $E$, \xmaxT and shower geometry while \rmuT can be derived using the measured signal in a station of the SD, defined as $S_\text{rec}$, by reversing \cref{eq:univ} and then fixing $S_\text{pred} = S_\text{rec}$ as in the following equation: 
\begin{equation}
    \rmuM = \frac{
        S_{\text{rec}} - \sum_{i} (1 - \alpha_{i})S_{i,\text{comp}}
    }{
        \sum_{i} \alpha_{i} S_{i}
    }.
\end{equation}
The terms $\alpha_i$ account for the correlation between the $i$-th component and the muonic 
component, where the coefficients $\alpha_\upmu = \alpha_{\text{e}\upgamma(\upmu)} = 1$.


\section{Dataset}
\label{sec::dataset}
This section details the dataset used for the search for UHE primary photons. 
The presented analysis is based on selected hybrid data
collected from 1 January 2005 to 31 December 2017. 

We adopted a \emph{blind analysis} approach, which consists in extracting a sub-sample of the data, corresponding to 5\% of the total and called \emph{burnt sample}, to define and optimize the analysis performance and to study the
expected background for this analysis. 
The hybrid dataset effectively used in the search for photons, after the subtraction of the burnt sample, amounts to about 2.8 million events. 
\subsection{Hybrid data selection}

Several selection criteria are applied to ensure a good quality and optimal resolution on reconstructed shower parameters, such as geometry, $E$ and \xmaxT. 
The event selection is inspired by the strategy adopted for previous hybrid analyses, both for mass composition measurements~\cite{PierreAuger:2014sui} and for photon searches with hybrid events~\cite{PierreAugerPh:2009, PierreAugerPh:2017, *PierreAugerPh:2017_erratum, PierreAuger:2022uwd}. It 
is carried out through four levels: pre-selection, geometry, longitudinal profile and quality of the atmosphere. 

\paragraph{Pre-selection:} The initial dataset consists of all events passing the conservative trigger requirements implemented in the data acquisition~\cite{PierreAuger:2015eyc}. Consequently, it still includes events to be removed for this analysis (e.g., lightning or low-energy events with a random-coincidence station). 
Events are rejected if the reconstruction process failed or if they have been recorded during time periods with known detection system issues (e.g., problems with the communications system or with unstable photomultipliers) or eventually without good FD or SD working conditions mostly occurring during the construction phase of the Observatory.

 \paragraph{Geometry:} To ensure that the probability of a trigger from at least one SD station is unity above 1 EeV regardless of the primary particle type, it is
required that the station selected in the hybrid reconstruction 
is  within \SI{1500}{\m} from the shower axis. The angular track length, defined as the angular separation between the highest and lowest elevation FD pixels in the track, has to be larger than \SI{15}{\degree}.
Events are selected if they land within a maximum distance from the triggered telescope such that the WCD trigger efficiency remains flat within 5\%~\cite{PierreAuger:2010swb,PierreAugerPh:2017,*PierreAugerPh:2017_erratum} when shifting the energy scale by its systematic uncertainty, i.e.\ by $\pm14\%$~\cite{spectrumPRD}. 
This distance, parameterized in different energy intervals, is based on simulations and is mostly
independent of the mass composition and hadronic models. 
Only events with zenith angle up to \SI{60}{\degree} are considered for this analysis.
More inclined events are not included because the absorption
of the electromagnetic components of the air shower in the atmosphere would be too high and the resultant 
trigger efficiency for photons too small in particular at the lowest energies.
As a consequence of these geometrical selection criteria, a resolution of about \SI{40}{\m} in the reconstructed core position and of
\SI{0.5}{\degree} in the determination of the arrival direction are reached for events with energy above 1 EeV.

\paragraph{Longitudinal profile:} the accuracy in the measurement of the longitudinal profile of a shower affects the resolutions on the reconstructed 
energy and depth of the shower maximum, \xmaxT.
A viewing angle between the shower axis and the telescope larger than \SI{20}{\degree} is required for rejecting events pointing towards the FD which have a large Cherenkov-light contamination. 
Biases in the reconstruction of the longitudinal profile are reduced by requiring that the observed \xmaxT is in the telescope field of view  and the fraction of gaps in the profile is 
smaller than 
20\% of the total observed length.
Moreover 
only events with a relative uncertainty on the calorimetric energy smaller than 20\% are accepted. These criteria ensure a resolution of the calorimetric energy at a level of 7 to 8\% 
and a resolution of \xmaxT 
below 20\,g~cm$^{-2}$, in line with the standard Auger analyses~\cite{DawsonICRC19,PierreAuger:2014sui}.

\paragraph{Quality of the atmosphere:} 
To exclude reflections or obscuration of fluorescence light by clouds, it is necessary to operate in a clean atmosphere, according to the combined information provided by several monitoring devices installed at the Observatory~\cite{PierreAuger:2012,icrc2019_violet}.
In particular, events are rejected if either the sky projection provided by the infrared cloud cameras or the ground-level projection provided by the GOES satellites indicates the presence of clouds over the array. 
When no data from these monitoring systems are available, the event is accepted only if during the data taking the average cloud fraction reported by Lidars operating at the FD sites is below 25\%. 
Finally, time periods  with poor viewing conditions are excluded, requiring that the vertical aerosol optical depth (VAOD), measured by the central laser facilities, integrated from the ground to \SI{3}{\km} is smaller than \num{0.1}.

\subsection{Universality and SD data selection}

The minimum requirement for accepting a signal from an SD station participating in a selected hybrid event is assessed using simulations.
The simulated showers have been produced using CORSIKA~\cite{CORSIKA} version 7.64, with  EPOS-LHC~\cite{WERNER200881} as the model for the description of the hadronic interactions at the highest energies, and FLUKA~\cite{Ferrari:2005zk} at lower energies. 
The showers are generated in the energy and zenith-angle ranges of interest for this analysis, i.e., between $10^{17.5}$\,eV and $10^{19.5}$\,eV, with angles between \SI{0}{\degree} and \SI{65}{\degree}.
The overall simulated data sample  consists of about 3 (6) million proton (photon)-initiated events. 
The simulation and reconstruction pipeline is based on the  
$\overline{{\rm Off}}\underline{{\rm line}}$
 software~\cite{offline} which combines a
detailed simulation of the FD and light propagation through the atmosphere 
with a GEANT4-based~\cite{geant4} simulation of the SD.
The detector response is reproduced accounting for the time-dependent configuration of the Observatory, that is, considering the actual status of the FD and SD data acquisition and the measured conditions of the atmosphere as a function of time, following the approach described in~\cite{PierreAuger:2010swb}.

\begin{figure}
\centering
\includegraphics[width=0.96\columnwidth]{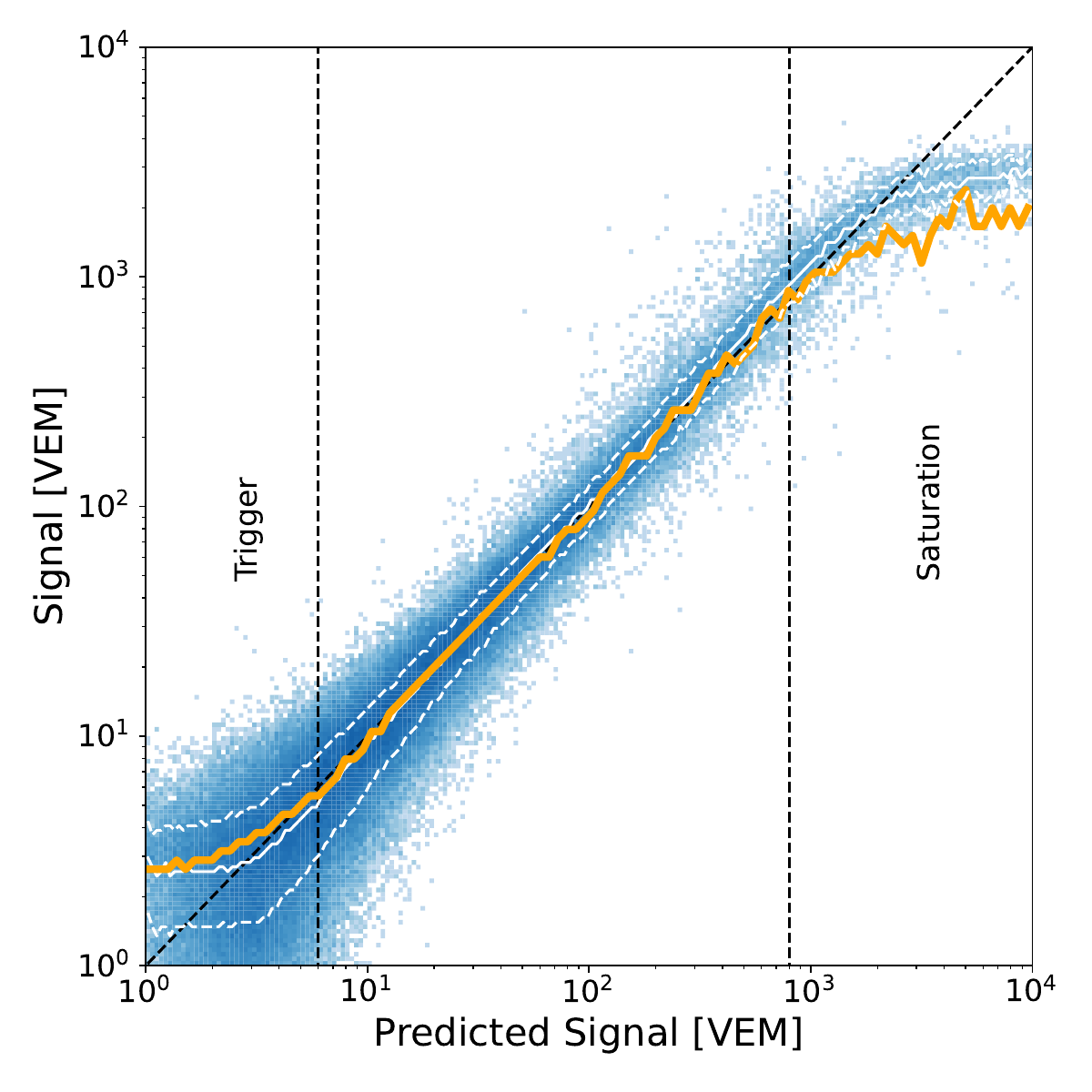}
\caption{Correlation between reconstructed and predicted signal for EPOS-LHC proton simulations (blue dots) with energies in the range of  
    $10^{17.5}$\,eV to $10^{19.5}$\,eV.
    Trigger effects are visible on the left, and saturation effects on the right, both extending beyond the region enclosed by the dashed lines. 
    Hybrid data from the burnt sample are shown on top of simulations (solid orange line). 
    }
\label{fig:signal_validation}
\end{figure}
As shown in \cref{fig:signal_validation}, the selection criteria for accepting signals in the SD stations are derived by studying the correlation between the reconstructed and predicted signal, for  simulations and for hybrid data.
Between 6 and 800\,VEM, the accuracy of the parametrization is better than 10\%. 
Biases appear below
\SI{6}{\VEM},\footnote{The signals reconstructed in the SD stations are measured in units of the signal produced by a vertical muon traversing the detector (VEM).} due to trigger effects, and above \SI{800}{\VEM} because of saturation effects. 
Thus, stations are included in the analysis if their signal size is in the range between 6 and \SI{800}{\VEM}.
Hybrid data from the burnt sample (solid orange line) agree with expectations at the level of 10\%  within the selected region. 

\begin{figure}
\centering
\includegraphics[width=0.95\columnwidth]{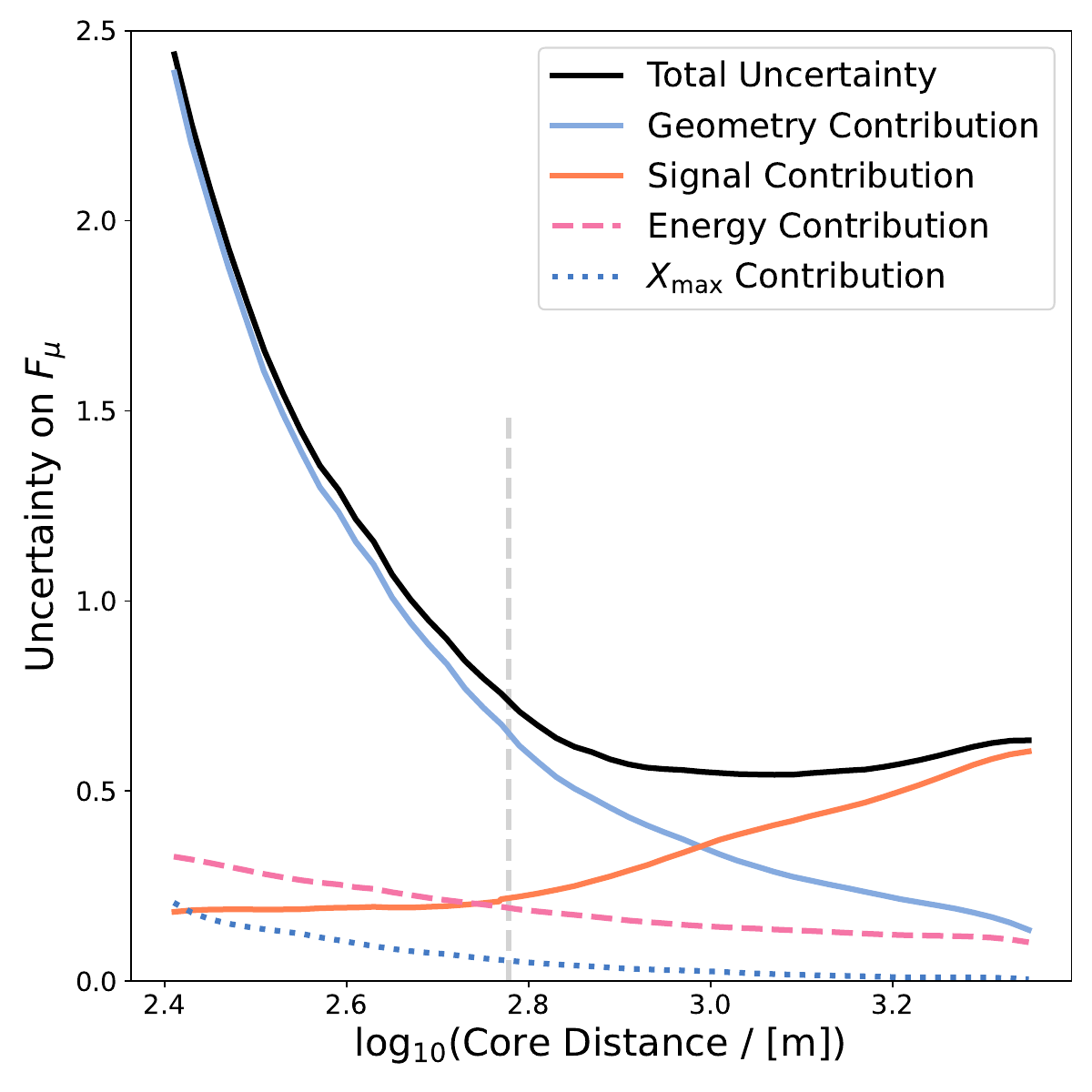}
\caption{Uncertainty on $F_\upmu$, $\sigma_{F_\upmu}$, (black solid line) as a function of the distance to the shower core.
    The colored lines show the different contributions:
    signal (orange, solid),
    geometry (blue, solid),
    energy (magenta, dashed) and
    \xmaxT (blue, dotted).
    The dashed gray vertical line marks the lower edge of the allowed core distance for this analysis (600 m).}
\label{fig:fmuErrors}
\end{figure}

The performance of the \rmuT calculation is studied as a function of the distance from the shower core by using proton-initiated showers. In particular, as shown in~\cref{fig:fmuErrors}, the overall uncertainty in \rmuT at distances less that \SI{600}{\m} (set as lower edge for this analysis) becomes very large 
and is dominated by the contribution due to the resolution of the geometric reconstruction in combination with the steepness of the air shower lateral distribution close to the axis. 
Moreover, above \SI{600}{\m} the overall uncertainty on \rmuT stabilizes within 15\% of its value up to the highest distances. 
The additional contributions to the overall uncertainty in \rmuT (also shown in~\cref{fig:fmuErrors})
are due, on the one hand, to the fluctuations of the signal and, on the other hand, to the reconstruction of the hybrid observables, such as energy and \xmaxT. 
When more than one station pass the described selection criteria for a given event, an average value of \rmuT is calculated by 
using the individual uncertainty as a weight. 

\begin{figure*}[htbp]
\centering
\def\w{0.9}
\subfloat[\label{fig:correlations:a}]{
  \includegraphics[width=\w\columnwidth]{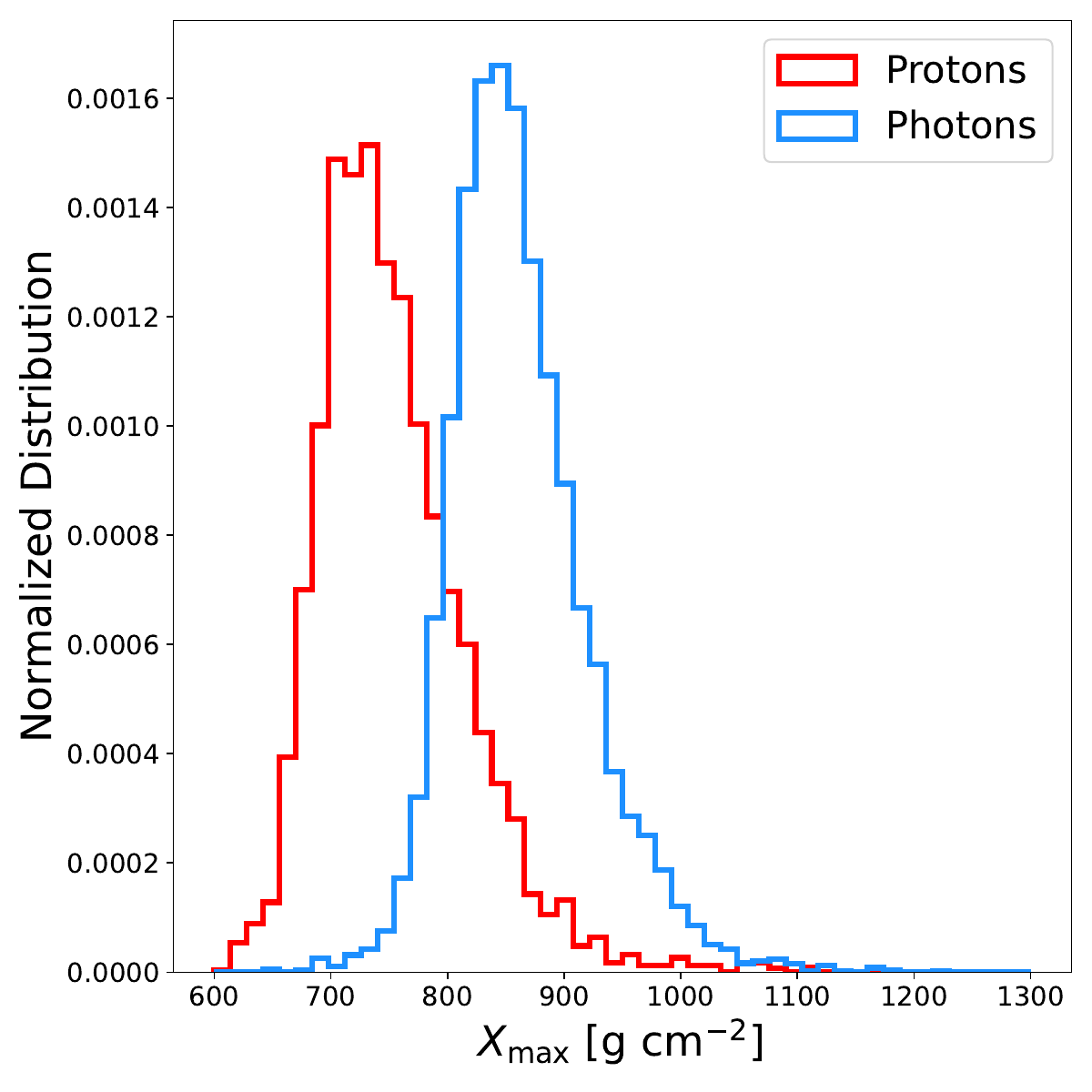}
}
\hfil
\subfloat[\label{fig:correlations:b}]{
  \includegraphics[width=\w\columnwidth]{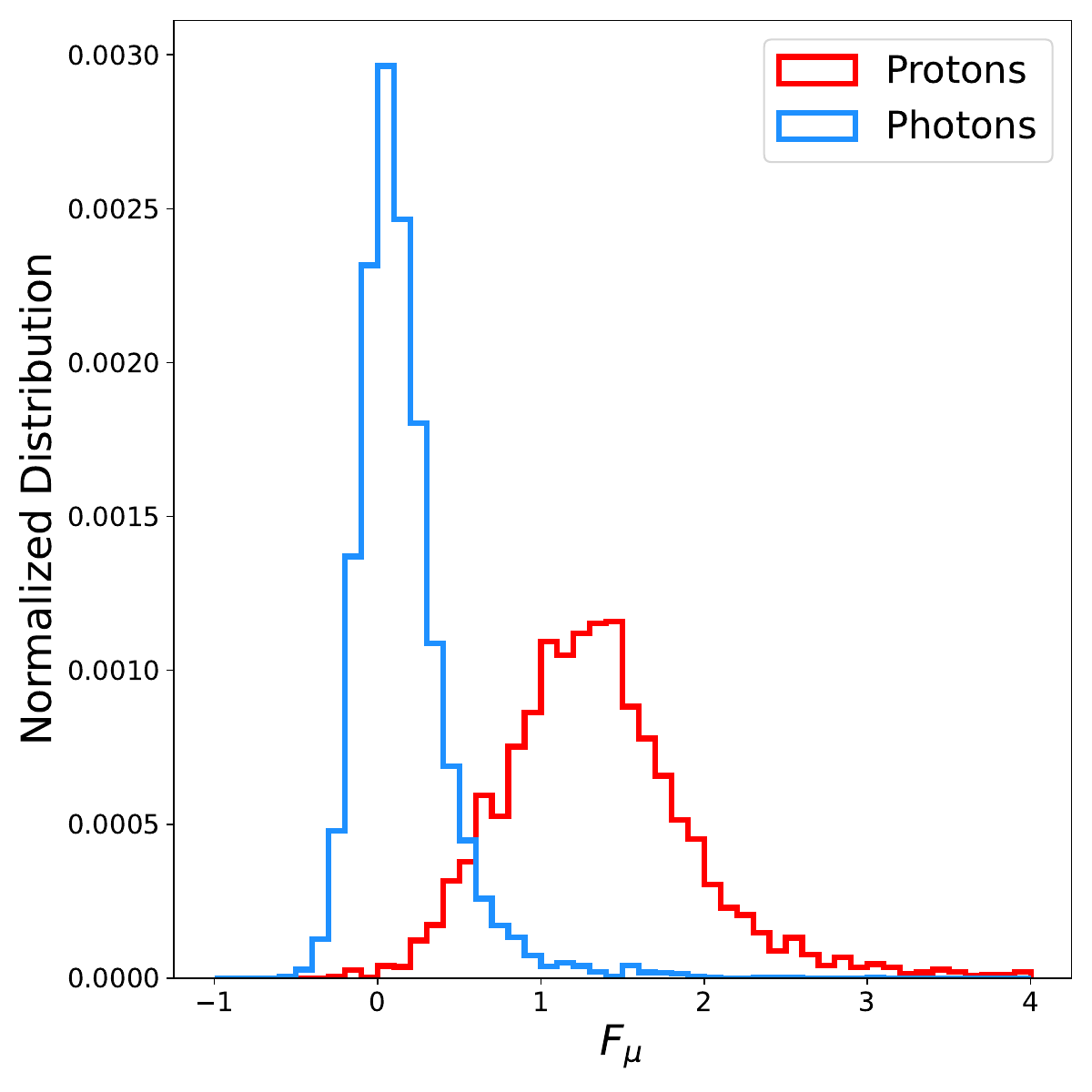}
}
\\    
\subfloat[\label{fig:correlations:c}]{
  \includegraphics[width=\w\columnwidth]{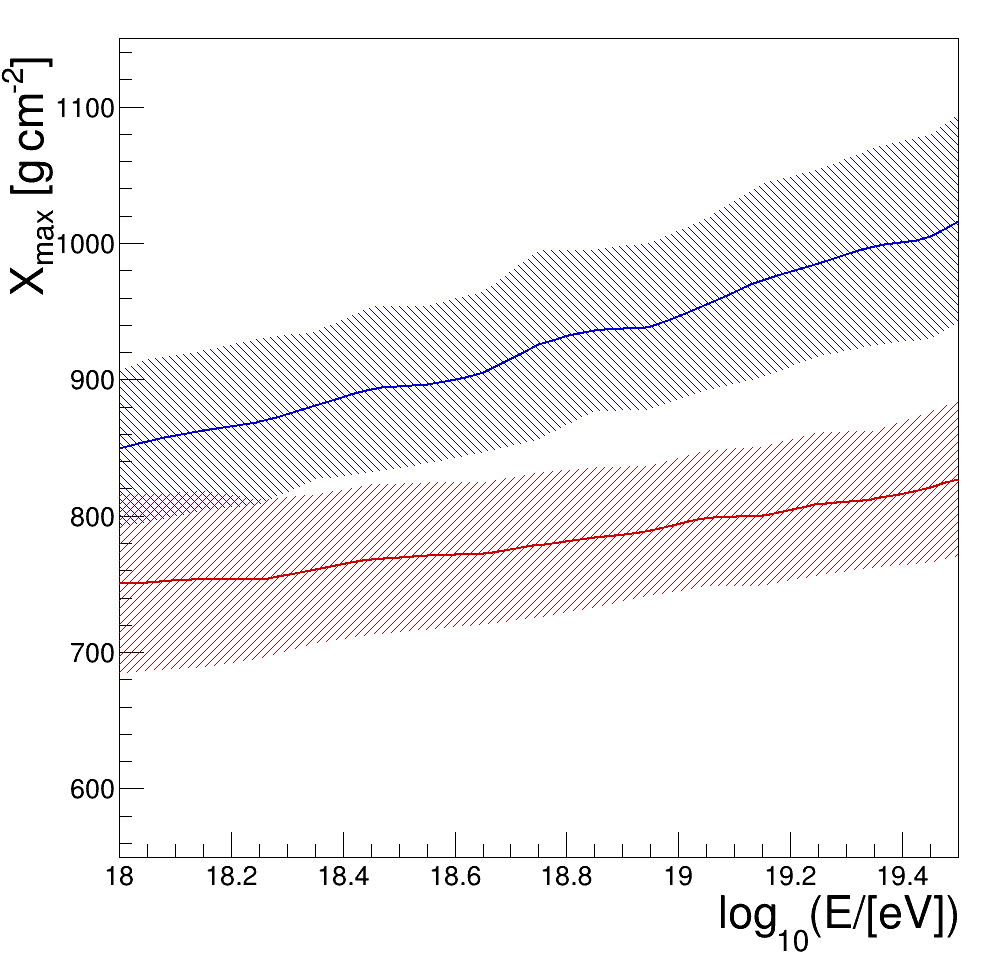}
}
\hfil
\subfloat[\label{fig:correlations:d}]{
  \includegraphics[width=\w\columnwidth]{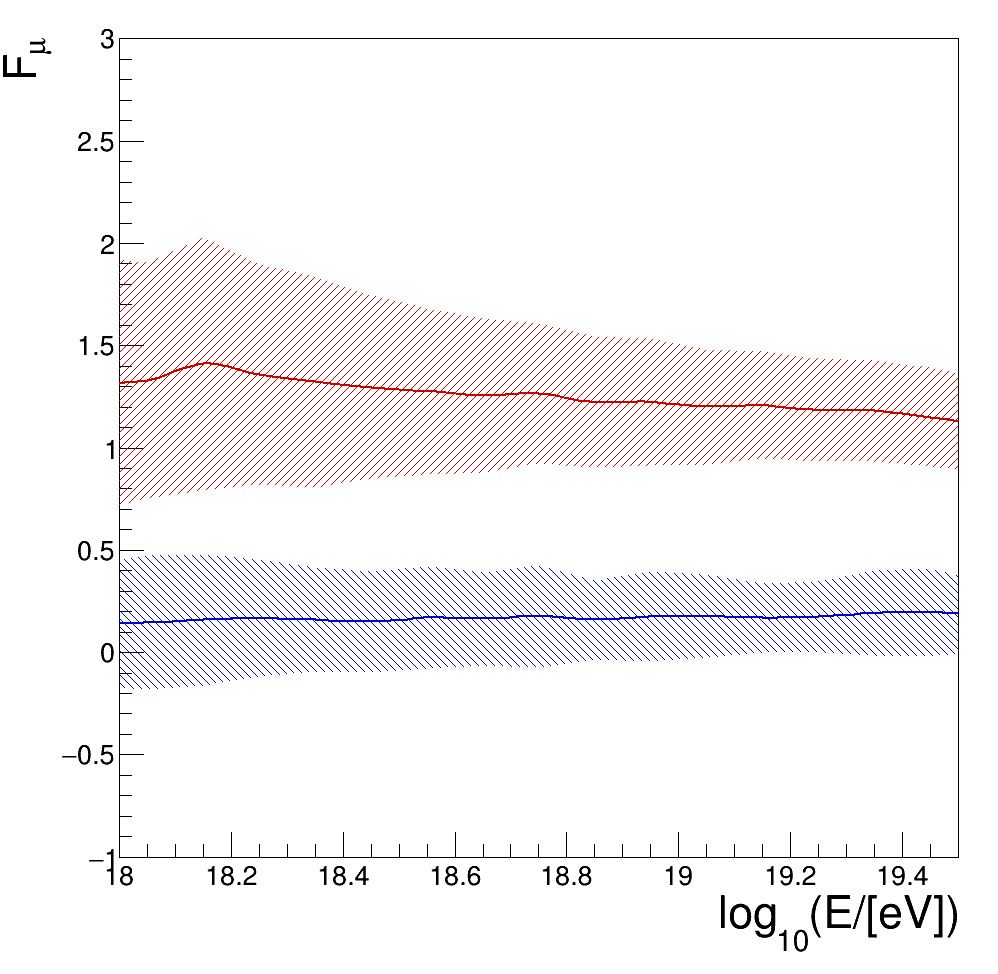}
}
\caption{Distributions of the reconstructed \xmaxT \textbf{(a)} and \rmuT \textbf{(b)} for photon-initiated (blue) and proton-initiated (red) air showers, with energies ranging between  $10^{18}$ eV and $10^{18.5}$ eV.
    The evolution of the average \xmaxT and \rmuT with respect to energy are shown in panel \textbf{(c)}  and \textbf{(d)} for protons (red) and photons (blue). 
    The shaded areas enclose one standard deviation. 
  }
\label{fig:correlations}
\end{figure*}

\begin{table}
\centering
\begin{tabular}{lrr}
\toprule
    & \multicolumn{1}{c}{$N$} & \multicolumn{1}{c}{\tabEffNam} \\
\midrule
    Raw Sample & 2\,990\,303 & \\
    Pre-selection & 1\,103\,316 & 36.9 \\
    Geometry & 393\,651 & 35.7 \\
    Profile & 198\,933 & 50.5 \\
    Atmosphere & 133\,741 & 67.2 \\
    Universality & 68\,886 & 51.5 \\
\bottomrule
\end{tabular}
\caption{Hybrid data: event selection criteria, number of events at different selection levels with the cut efficiency $\epsilon$, calculated with respect to the preceding cut. See text for details on the definition of the selection levels.}
\label{11::tab::cuts::selectionEfficiency}
\end{table}

\cref{11::tab::cuts::selectionEfficiency} shows the effect of the selection criteria on the data.
Overall, out of the 2\,990\,303 events in the full dataset, 68\,886 events are selected while 
in the simulation samples, the same selection pipeline yields 17\,215 proton-initiated events and 22\,237 photon-initiated ones.


\section{Combining FD and SD observables: the MVA approach}
\label{sec::combination}

\begin{figure*}[ht!]
\centering
\def\w{0.8}
\subfloat[\label{fig:combination:a}]{
  \includegraphics[width=\w\columnwidth]{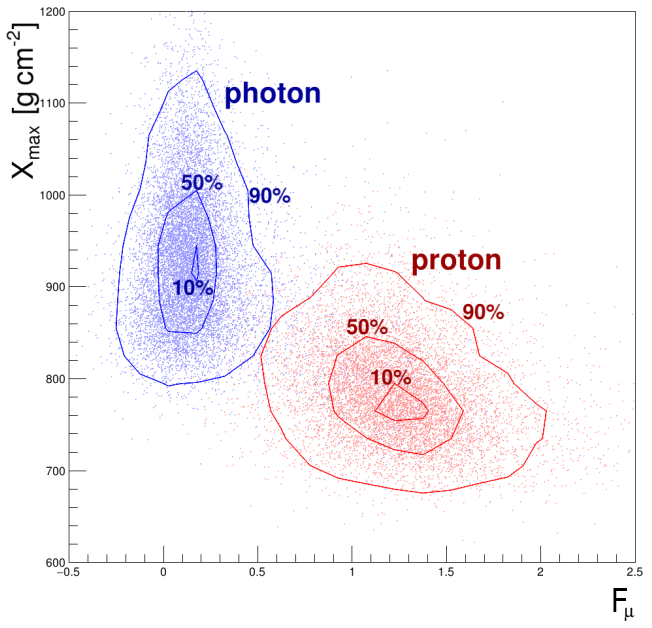}
}
\hfil
\subfloat[\label{fig:combination:b}]{
  \includegraphics[width=\w\columnwidth]{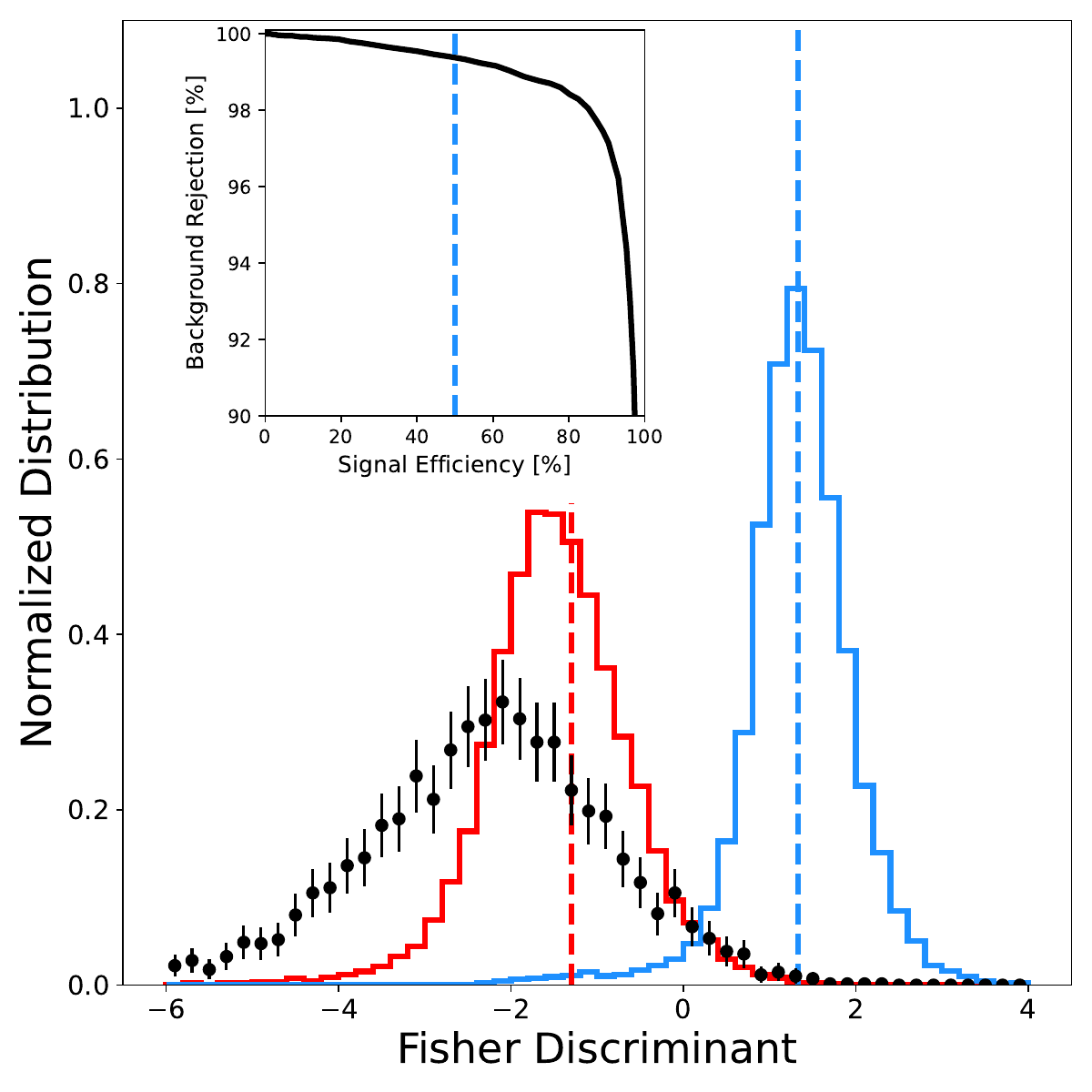}
}
\caption{%
    \textbf{(a)}: \xmaxT-\rmuT distributions for photons (blue) and protons (red).
    Contour lines enclose the 90\%,
    50\% and 10\% of the distributions of the events,
    re-weighted to a realistic power law spectrum $E^{-\Gamma}$ ($\Gamma=2.7$ for protons
    and $\Gamma=2.0$ for photons)~\cite{SavinaICRC21}.
    \textbf{(b)}: Distribution of the Fisher discriminant for simulated photons (signal, blue) and protons (background, red), 
    and for the burnt sample (black). 
    The vertical red line indicates the value of the Fisher discriminant (\fisherT=$-1.3$) above which the background begins to decrease nearly exponentially, while the photon selection efficiency is still very close to 100\%; the blue line indicates the median of the photon distribution.
        \textbf{(Inset)}: background rejection as a function of signal efficiency obtained with the Fisher Discriminant Analysis.
}
\label{fig:combination}
\end{figure*}

The selection criteria discussed in \cref{sec::dataset} ensure a good resolution of the two key variables for photon/hadron separation, 
\xmaxT and \rmuT. 
Their distributions are shown in panels (a) and (b) of \cref{fig:correlations}, respectively, for photon (signal) and proton (background) initiated showers within the energy range from $10^{18.0}$\,eV to $10^{18.5}$\,eV.
These simulations are re-weighted assuming a power-law energy distribution  
\mbox{ 
d$N$/d$E$ $\propto$ E$^{-\Gamma}$} with spectral index $\Gamma=2.7$ for protons and $\Gamma=2.0$ for photons, motivated by previous Auger photon searches~\cite{PierreAugerPhTar:2014, *PierreAugerPhTar:2017, PierreAugerPh:2017, *PierreAugerPh:2017_erratum} . 
Panels (c) and (d) illustrate the dependence of \xmaxT and \rmuT on energy.
Overall, the discussed variables provide an excellent separation power between the considered primary species.
While the separation power for this analysis is comparable to that used in~\cite{PierreAugerPh:2017, *PierreAugerPh:2017_erratum},
the $\rmuT$-based criterion provides a selection efficiency that improves with energy, exceeding that used in \cite{PierreAugerPh:2017, *PierreAugerPh:2017_erratum} above 2.5 EeV and reaching 100\% at the highest energies.
Furthermore, \rmuT is almost independent of the primary particle energy, whereas \xmaxT shows a logarithmic dependence on the energy.

The performance of the combined observables in terms of photon/hadron separation is expected to surpass that of each variable separately, as illustrated in panel (a) of \cref{fig:combination}. The blue (red) contour lines enclose 90\%, 50\% and 10\% of the photon (proton) distributions. They have clearly separated peaks, with minimal overlapping tails.
The \rmuT and \xmaxT parameters do not show any significant degree of correlation. 

To maximize the potential of the hybrid approach, \rmuT is then combined with \xmaxT within the framework of a multivariate analysis (MVA). 
The dependence of the two variables on the energy is also taken into account as an additional parameter. 
In the following, the variable defined as 
$E_\upgamma = (1 + 1\%)\ E_\text{cal}$ will be used 
as an estimator for the primary (photon) energy. $E_\text{cal}$ is the calorimetric 
shower energy reconstructed by the FD and the factor 1\% accounts for the invisible energy contribution due to neutrinos and high-energy muons ending underground~\cite{PierogEngelHeck2006}. 
$E_\upgamma$ will be used as default for simulations and data, independently of the nature of the primary particle.


More specifically, a Fisher discriminant analysis~\cite{Fisher36} was performed, using  three input parameters, \rmuT, \xmaxT and $\log_{10}(E_\upgamma)$ combined linearly to obtain a Fisher discriminant, \fisherT.
The event classification is then made in the transformed \fisherT space. 
The linear discriminant analysis identifies an axis in the hyperspace of the input variables such that, when projecting the output classes (signal and background) along this axis, the separation between the two classes is maximized, while the dispersion of the simulated events within each class is minimized. 
The use of a Fisher discriminant analysis is an appropriate choice for this case as it provides a robust event classification for uncorrelated input observables, which is the case for \rmuT and \xmaxT. In addition to that, the discriminant can be calculated analytically for each event.  
In \cref{fig:combination} (b), the distribution of the Fisher discriminant is shown for simulated photons (blue) and protons (red), along with data from the burnt sample (black). 
The vertical red line indicates the value of the Fisher discriminant ($\fisherM=-1.3$) above which the background begins to decrease nearly exponentially, while the photon selection efficiency is still very close to 100\%; the blue line indicates the median of the photon distribution.
Assuming photons as signal and proton as background, the background rejection power as a function of signal efficiency is shown in the inset panel of \cref{fig:combination} (b).
As an example, for a signal efficiency of 50\% (dashed blue line), a background rejection at the level of 99.75\% is achieved. 



\section{Background Expectation}
\label{sec::background}

\begin{figure}
\centering
\includegraphics[width=\columnwidth]{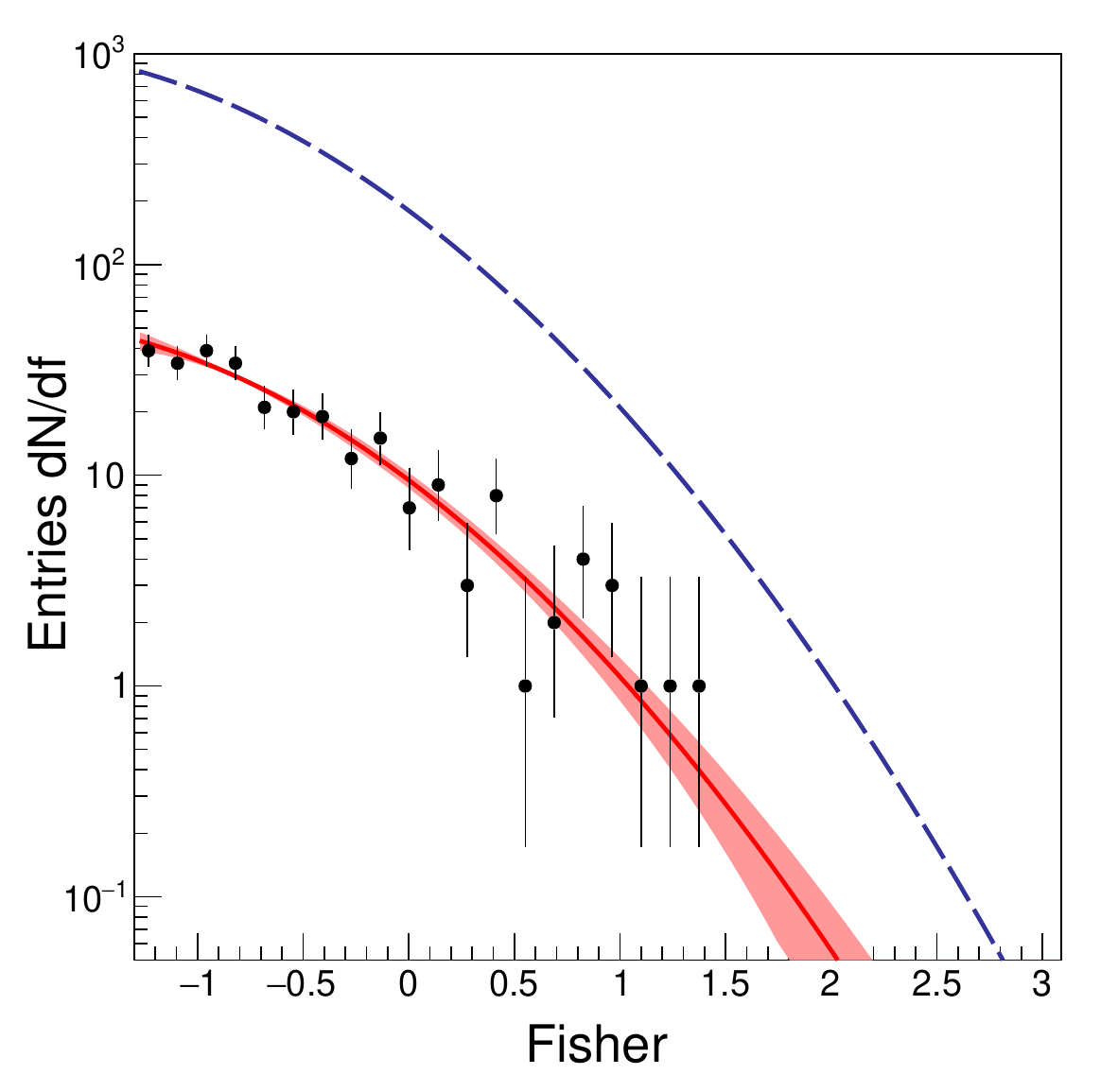}
\caption{Distribution of the Fisher discriminant 
  for the burnt sample (black points), together with the fit to the $m$ function (red line) and the 1$\sigma$ uncertainty (red band). 
  The dashed blue line shows the background expected in the full hybrid data sample.}
\label{fig:background}
\end{figure}

For this analysis, the expected amount of background is calculated from data. 
Firstly, the distribution of the Fisher discriminant \fisherT has been parameterised above the value \fisherT=$-1.3$ introduced in section~\ref{sec::combination}.
The shape of the distribution has been modeled based on proton simulations, assuming an exponential function $m$ defined as
%
\begin{equation}
\label{07::eq::model::background}
  m(\fisherM |A,B) = N(A,B) e^{-(A\fisherM^{2}+B\fisherM)},
\end{equation}
where $A$ and $B$ are shape parameters and $N$ is a normalization factor, calculated by requiring that the integral of $m$ above $-1.3$ equals the number of events in the burnt sample in the same range of \fisherT. The fit of the Fisher distribution to the burnt data is shown in \cref{fig:background} as a red line superimposed to data points. 




A possible photon contamination in the burnt sample cannot be excluded: a related systematic effect has thus been studied by using a jackknife technique~\cite{EfronStein}.
This is a re-sampling technique, which involves a leave-one-out strategy for the estimation of the parameters (in this case, $A$ and $B$) in a data set of $n$ observations.
A detailed description of the background estimate, along with the values of the parameters $A$ and $B$ and the corresponding uncertainties 
is given in the~\cref{sec::appendixA}.

%

The distribution of the Fisher discriminant for the extrapolated background is shown as a blue line in
\cref{fig:background} along with its $1\sigma$ uncertainties included as a blue band.   


The extrapolation of the background is thus used to determine the optimal value of the Fisher discriminant cut, $\fisherM_\upgamma$, 
for selecting photon candidates.
The upper limit values can be optimized by following the approach described in ~\cite{Savina:2021xpt}, in the case of signal non-observation. In this way, 
$\fisherM_\upgamma$ has been found to be approximately at the median of the Fisher discriminant distribution for photons. 
The median value ($\fisherM_\upgamma$ ${\simeq}1.4$, blue vertical line in \cref{fig:combination}), has been adopted as photon candidate cut.   
Finally, the number of expected false-positive events in the full hybrid data set can be calculated by integrating the function describing the extrapolated background above $\fisherM_\upgamma$, and it yields \num{30(15)}.


\section{Photon Search in data}
\label{sec::results}
The method is applied to the full hybrid data sample that, after the application of the selection criteria described in \cref{sec::dataset}, results in 68\,886 hybrid events of which 26\,752 have E$_{\gamma}$ above 10$^{18}$ eV.

\begin{figure*}
\centering
\subfloat[\label{fig:unblinding:a}]{
  \includegraphics[width=0.95\columnwidth]{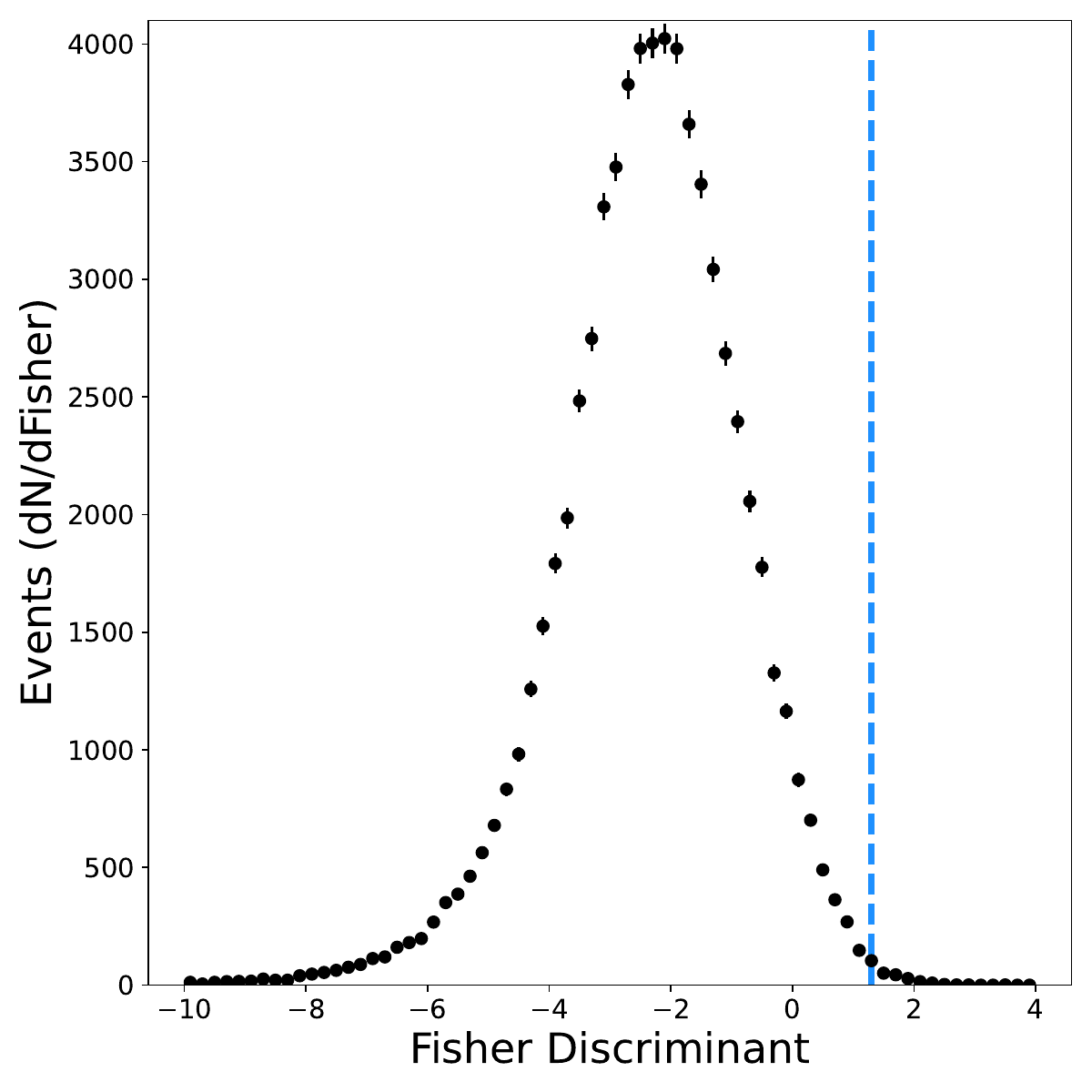}
}
\hfil
\subfloat[\label{fig:unblinding:b}]{
  \includegraphics[width=\columnwidth]{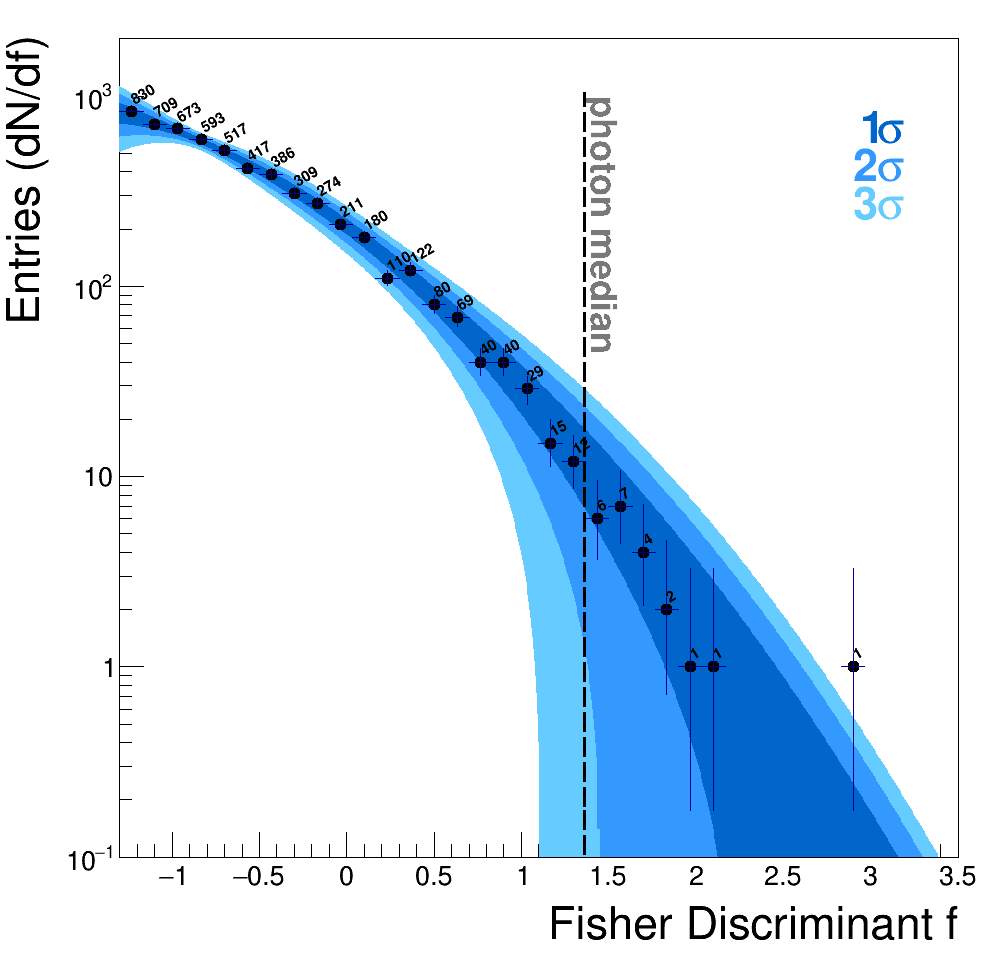}
}
\caption{\textbf{a:} Fisher discriminant distribution of the selected hybrid data sample;
\textbf{b:} Tail of the Fisher discriminant distribution ($\fisherM > -1.3$) of the hybrid data sample (black dots).
The vertical dashed line represented the photon-median cut.
The shaded blue areas show the 1, 2, $3\sigma$ uncertainties in the expected background.}
\label{fig:unblinding}
\end{figure*}


\begin{figure}
\centering  
\includegraphics[width=0.95\columnwidth]{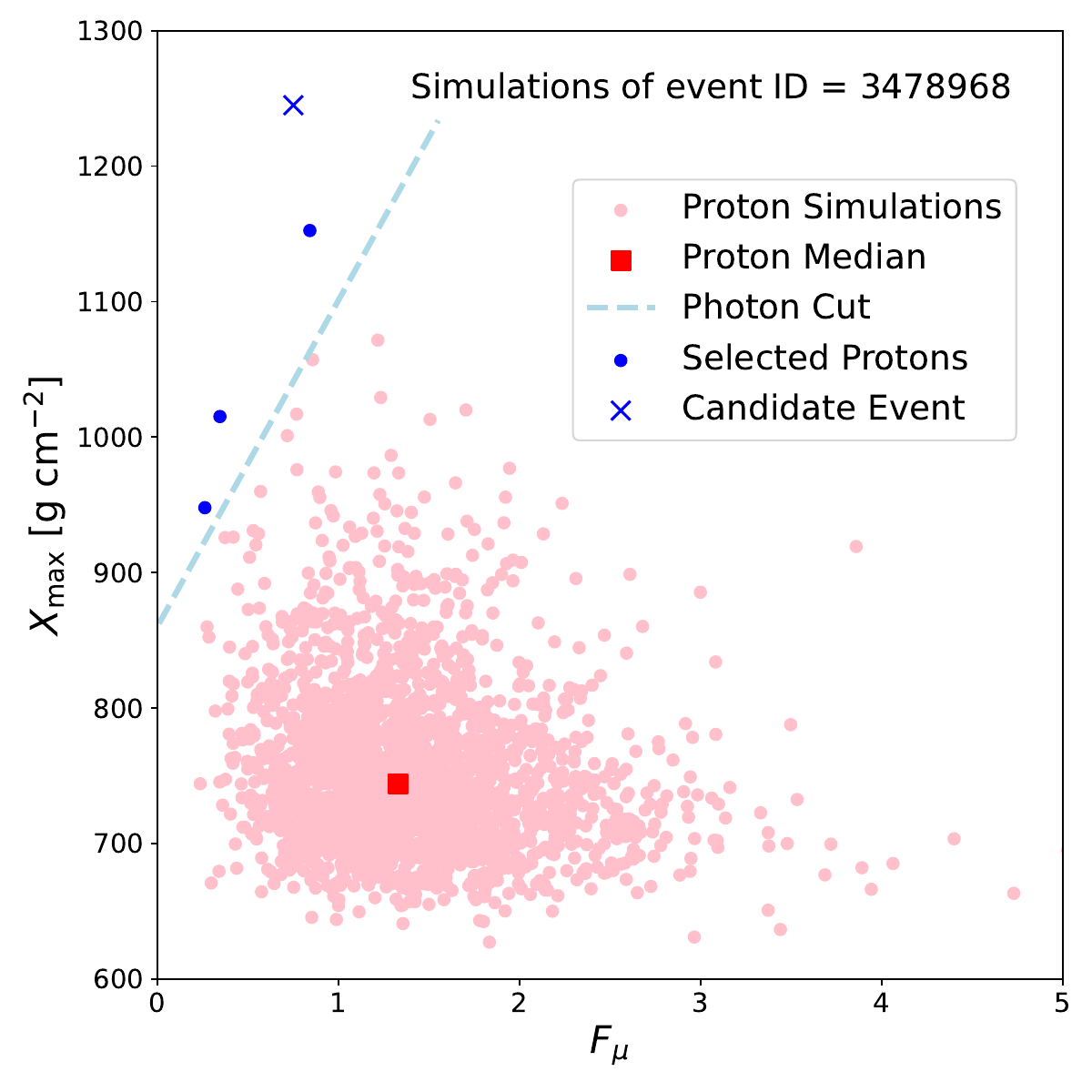}
\caption{
Simulation of 2000 proton showers with the same energy, geometry and detector configuration as the most significant candidate event with ID 3478968 (blue cross).
The reconstructed \xmaxT and \rmuT of the simulated events are shown as light red dots. 
The light blue dashed line shows the photon cut at the energy of the candidate event.
Three out of them are selected as false-positive photon candidates (blue circles).
}
\label{fig:lookElsewhere}
\end{figure}

\begin{figure}
\centering
\includegraphics[width=0.9\columnwidth]{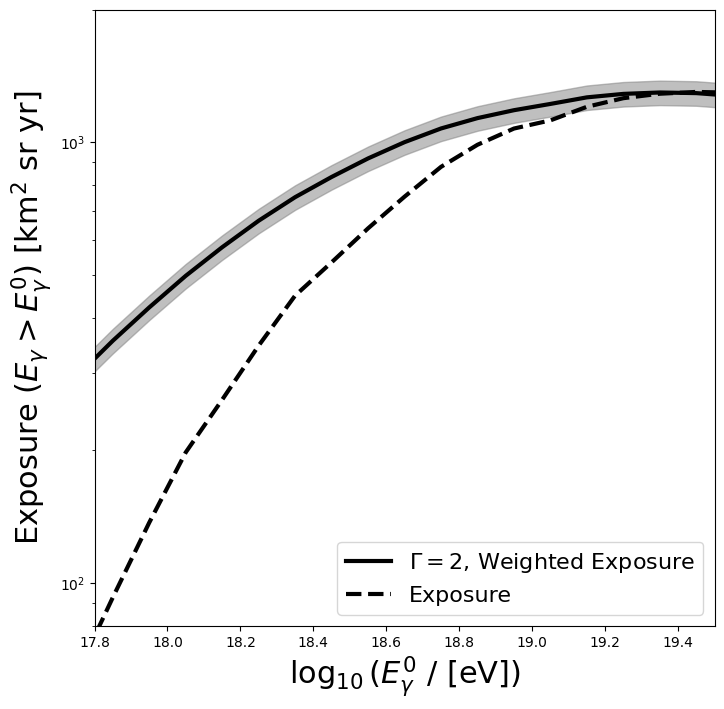}
\caption{Weighted hybrid exposure for primary photons (solid line) in the time interval from 1 January 2005 to 31 December 2017, assuming a power-law spectrum 
with $\Gamma$ = 2.
Systematic uncertainties due to the on-time and the trigger efficiency are shown as a gray band. The raw exposure (dashed line) is also shown for comparison.}
\label{fig:exposure}
\end{figure}

The distribution of \fisherM , obtained by combining \xmaxT and \rmuT in the dataset, is
displayed in \cref{fig:unblinding} (a) where the vertical dashed blue line represents 
the photon selection cut.
In \cref{fig:unblinding} (b) the tail of the Fisher distribution ($\fisherM > -1.3$),
enclosing ${\sim}5600$ events (black points), is shown along with
the shaded blue bands representing the expected background, 
with uncertainties at different sigma levels.

As one can see, the data distribution is compatible with that from the expected background.
The median selection cut yields 22 candidates, which is well-consistent with the expectation
of $30 \pm 15$ false-positive candidates, as calculated in the previous section.
The general characteristics of the candidate events are summarised in
\cref{08::tab::candidates::candidates},
where $E_\upgamma$, \xmaxT, \rmuT, the 
UTC time and the value of the Fisher
discriminant are reported for each candidate.

\begin{table*}
\centering
\begin{tabular}{rlccrrr}
\toprule
 & UTC time & $\lg(E_\upgamma/\text{eV})$ & \xmaxT/(g~cm$^{-2}$) & \hspace{0.3cm} \rmuT \hspace{0.1cm} &
 \hspace{0.5cm} $\theta/^\circ$ &\hspace{0.5cm} Fisher  \\
\midrule
Jun 22,& 2006 07:27:16 & 18.31 & 987.7 & 0.42 & 38.7 & 1.57 \\ 
Jun 27,& 2006 03:01:26 & 18.01 & 1039.9 & 0.39 & 47.6 & 2.12 \\ 
May 22,& 2007 02:58:14 & 18.24 & 1245.2 & 0.75 & 56.7 & 2.87 \\ 
Aug 10,& 2007 03:05:06 & 18.02 & 907.6 & 0.22 & 43.6 & 1.46 \\ 
Dec 15,& 2007 06:29:00 & 18.00 & 913.4 & 0.29 & 47.8 & 1.40 \\ 
Mar 26,& 2009 06:34:56 & 18.10 & 938.9 & 0.11 & 39.0 & 1.84 \\ 
Oct 19,& 2009 06:54:20 & 18.29 & 1008.7 & 0.52 & 47.8 & 1.57 \\ 
Oct 21,& 2009 03:51:13 & 18.01 & 1010.4 & 0.59 & 59.3 & 1.58 \\ 
Jan 19,& 2010 03:55:42 & 18.21 & 796.3 & -0.23 & 22.7 & 1.36 \\ 
Oct 03,& 2010 05:07:00 & 18.01 & 1019.9 & 0.52 & 49.6 & 1.75 \\ 
Oct 16,& 2010 07:33:46 & 18.14 & 984.7 & 0.45 & 47.3 & 1.57 \\ 
Jun 26,& 2011 05:17:41 & 18.17 & 935.6 & 0.07 & 30.8 & 1.86 \\ 
Jul 05,& 2011 06:17:13 & 18.02 & 1109.3 & 1.01 & 57.2 & 1.57 \\ 
Aug 03,& 2011 01:59:06 & 18.20 & 944.3 & 0.20 & 54.6 & 1.68 \\ 
Dec 22,& 2011 05:31:33 & 18.08 & 932.7 & 0.02 & 44.2 & 1.96 \\ 
Nov 13,& 2012 06:51:13 & 18.04 & 967.5 & 0.48 & 35.0 & 1.45 \\ 
Jun 30,& 2013 02:01:08 & 18.04 & 1061.8 & 0.86 & 41.7 & 1.47 \\ 
Mar 15,& 2015 06:32:28 & 18.48 & 1001.9 & 0.45 & 51.8 & 1.55 \\ 
Mar 08,& 2016 01:23:38 & 18.04 & 954.3 & 0.29 & 54.5 & 1.67 \\ 
Jul 05,& 2016 06:01:34 & 18.12 & 917.0 & 0.07 & 48.1 & 1.74 \\ 
Aug 11,& 2016 07:52:15 & 18.07 & 847.4 & 0.01 & 58.5 & 1.38 \\ 
Jun 19,& 2017 01:14:36 & 18.05 & 849.9 & -0.07 & 42.4 & 1.54 \\ 
\bottomrule
\end{tabular}
\caption{Details of the events selected by the photon candidate cut.}
\label{08::tab::candidates::candidates}
\end{table*}

The candidate event 
with the highest Fisher value has the peculiarity
of having a very deep \xmaxT. 
The event, labeled with the ID 3478968 was detected on 22 May 2007 at 02:58:14\,UTC. 
The atmospheric conditions at the time of the event were checked and found to be optimal,
with a measured VAOD of 0.02 and no cloud coverage.
The hybrid reconstruction yields an energy $E_\upgamma = (1.73 \pm 0.16){\times}10^{18}$\,eV,
a depth of the shower maximum $\xmaxM = (1245 \pm 57)$\,g~cm$^{-2}$ 
and a zenith angle $\theta=(56.7 \pm 1.0)^\circ$. The footprint of the event on the SD array 
is characterized by six triggered stations. 
Out of them, only three stations pass the selection criteria for the calculation of \rmuT described in  \cref{sec::dataset}.
The station with the largest signal size is, in fact, rejected because it is too close to the core, while the other two excluded stations are rejected because of their small signal size. 
The \rmuT associated with this event is $0.75 \pm 0.41$. By combining it with the value of \xmaxT, the resulting value of the Fisher discriminant is $\fisherM\simeq2.87$

The event was also cross-checked with SD-based information. Namely, the risetime of the signals in the triggered stations 
has been analysed and found to be consistent with an event developing late in the atmosphere.

To study more in detail this specific candidate, 2000 proton showers, characterised by the same geometric configuration and energy as the candidate, were simulated (see \cref{fig:lookElsewhere}) 
with CORSIKA using EPOS-LHC as the model for high-energy hadronic interaction. 
It turns out that only three proton events (blue dots) lie beyond the photon cut at the energy of the candidate but none out the 2000 ones exhibit a Fisher value larger than that of the candidate (blue cross), namely $f_c=2.87$.
The probability to observe a background event with the same zenith angle and energy that yields to a Fisher value larger than $f_c$ is consequently $<$ 1/2000. However, because \xmaxT and \rmuT
can combine to form a Fisher value larger than $f_c$ for showers whose zenith angle and energy lie anywhere in the parameter space explored in the analysis ("look-elsewhere" effect~\cite{Lyons2008}), this probability is only \textit{local}. 
To quantify the compatibility between the selected candidate and a background event \textit{globally}, we have then simulated a large number of showers to probe the entire parameter space. 100,000 realisations of the data sample using the extrapolated Fisher distribution have been generated as a background model. 
The p-value characterising the deviation of $f_c$ from the background model is obtained by counting the number of realisations with $max(f)>f_c$ out of the total. 
It amounts to 25\%, which gives a very modest overall significance for claiming that the candidate is a photon.





\section{Photon Flux upper limits}
\label{sec::discussion}
\begin{figure*}[hbt!]
\centering
\includegraphics[width=0.95\textwidth]{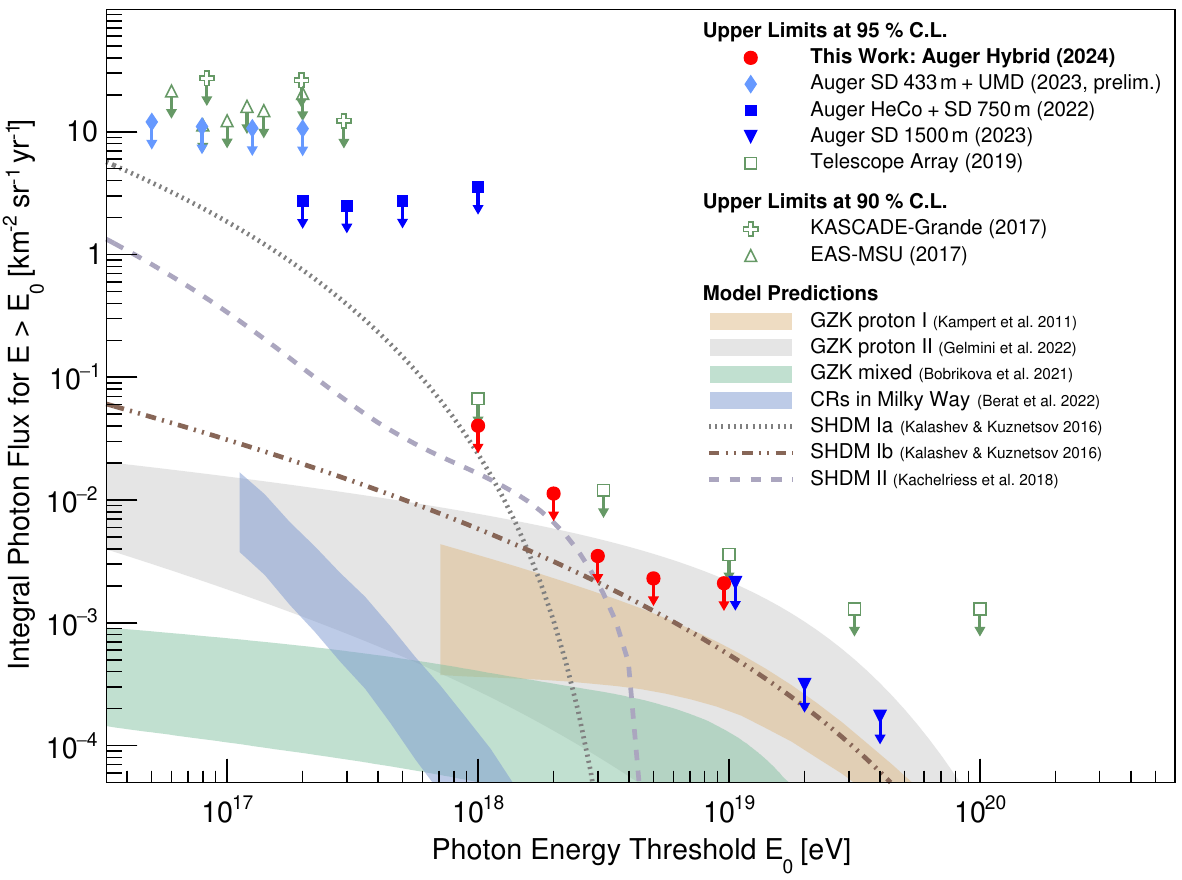}
\caption{
Current upper limits on the integral photon flux determined with data collected by the
Pierre Auger Observatory (solid markers) from ~\cite{PierreAuger:2022uwd, PierreAugerPh:2023} and preliminary limits from~\cite{GonzalezICRC23}. Additionally, we display upper limits reported by other experiments: KASCADE-Grande (green crosses)~\cite{Apel_2017}, EAS-MSU (green triangles)~\cite{PhysRevD.95.123011}, and Telescope Array surface detector (green squares)~\cite{ABBASI20198}. The ranges of expected
GZK photon fluxes under the assumption of two different pure-proton scenarios are depicted as the brown
and gray bands (adapted from~\cite{Sarkar:2011hkm} and~\cite{universe8080402}, respectively). The water green band illustrates the expected GZK photon flux, assuming a mixed composition that aligns with Auger data~\cite{Bobrikova:2021wG}, while the light blue band signifies
the range of photon fluxes expected from cosmic-ray interactions with matter in the Milky Way~\cite{Berat_2022}. Additionally, the expected photon fluxes from the decay of super-heavy dark matter particles for different masses and lifetimes are included~\cite{PhysRevD.94.063535, PhysRevD.98.083016}, see text for more details.
}
  \label{fig:PhotonLimits}
\end{figure*}

Since no significant excess of photons has been observed with respect to the background, upper limits on the diffuse UHE photon flux, $\Phi_\upgamma^\text{C.L.}$, are derived as
\begin{equation}
\label{08::eq::ul::ul}
  \Phi_\upgamma^\text{C.L.}(E_\upgamma{>}E_\upgamma^0) =
  \frac{N_\upgamma^\text{C.L.}(E_\upgamma{>}E_\upgamma^0)}
       {\calE_\upgamma(E_\upgamma{>}E_\upgamma^0)} 
\end{equation}
where $N_\upgamma^\text{C.L.}$ is the upper limit on the number of photons at a certain confidence level above an energy threshold $E_\upgamma^0$, and $\calE_\upgamma$ is the weighted hybrid exposure to photons above the same $E_\upgamma^0$.


\subsection{The hybrid photon exposure}
\label{sec::exposure}
The raw exposure of the hybrid detector to photons is calculated following the approach detailed in~\cite{PierreAuger:2010swb}, as
\begin{equation}
\label{08::eq::ul::aperture}
  \calE_\upgamma = 
  \int\mathrm{d}t\int\mathrm{d}\Omega\cos\theta\int\mathrm{d}S \, \varepsilon_\upgamma(E_\upgamma, t, \theta, \phi, x, y),
\end{equation}
where $\varepsilon_\upgamma$ is the overall photon efficiency, including detection, reconstruction, and selection of events.
$\varepsilon_\upgamma$ is a function of zenith angle $\theta$, azimuth angle
$\phi$, impact position $x,y$, time $t$ and energy,
$E_\upgamma$.

The raw exposure $\calE_\upgamma$ is then weighted with a power
law spectrum assuming a spectral index $\Gamma=2$, namely
\begin{equation}
\label{07::eq::exposure::weighted}
  \calE_\upgamma^\text{weighted}(E_\upgamma{>}E_\upgamma^0) =
  \frac{\int_{E_\upgamma^0}^\infty E_\upgamma^{-\Gamma}
        \calE_\upgamma(E'_\upgamma)\,\mathrm{d}E_\upgamma}
       {\int_{E_\upgamma^0}^\infty E_\upgamma^{-\Gamma}\,\mathrm{d}E_\upgamma}.
\end{equation}
The resulting behavior of $\calE_\upgamma^\text{weighted}$ as a function of the energy threshold is shown as a solid line in \cref{fig:exposure}. The grey shaded band represents the contribution to the systematic uncertainty due to on-time and trigger efficiency, estimated at the level of $\pm6.4\%$~\cite{PierreAugerPh:2017,*PierreAugerPh:2017_erratum}. 

\subsection{Upper limits}
\label{08::sec::ulCalc}

The calculation of upper limits is carried out through \cref{08::eq::ul::ul} for
five different energy thresholds, $E_\upgamma^0$, the same as in~\cite{PierreAugerPh:2017,*PierreAugerPh:2017_erratum}.
They are listed in the first column of \cref{08::tab::ul::nBackground2}.

The values of $N^{95\%}_\upgamma$  (shown in the fifth column of the table) are computed for each specified threshold as the Rolke~\cite{Rolke2005} upper limit at 95\%\,C.L.\ of the observed number of photon candidates listed in the fourth column. In this calculation, 
the expected number of background events and its largest uncertainty, reported in the second and third  columns respectively, is derived from the parameterisation of the background given in \cref{sec::background}, after normalizing it to the number of hybrid events above each $E_\upgamma^0$.
The weighted exposure $\calE_\upgamma^\text{weighted}$ is displayed in the sixth 
  column.

The integral flux upper limits can also be converted into photon fractions upper limits, relative to the measured cosmic rays flux~\cite{spectrumPRD}, thus leading to 0.15\%, 0.21\%, 0.15\%, 0.26\%, 0.77\% above the corresponding energy thresholds given in the first column of \cref{08::tab::ul::nBackground2}.


\begin{table*}[hbt!]
\centering
\begin{tabular}{cccccc}
\toprule
    $E_\upgamma^0\qquad$
    & $N_\text{b}(E_\upgamma{>}E_\upgamma^0)$
    & $\quad N_\upgamma(E_\upgamma{>}E_\upgamma^0)$
    & $\quad N^{95\%}_\upgamma(E_\upgamma{>}E_\upgamma^0)$
    & $\quad \calE_\upgamma^\text{weighted}(E_\upgamma{>}E_\upgamma^0)$
    & $\qquad \Phi^{95\%}_\upgamma(E_\upgamma{>}E_\upgamma^0)$
\\
  (EeV)
  &
  & 
  &
  & (km$^2$\,sr\,yr)
  & (km$^{-2}$\,sr$^{-1}$\,yr$^{-1}$)
\\
\midrule \vspace{5pt}
    1  & $30^{+15}_{-15}$        & 22 &  23.38 & 579  & 0.0403 \\\vspace{5pt}
    2   & $6^{+6}_{-2}$       & 2  &   9.53 & 840  & 0.0113 \\\vspace{5pt}
    3   & $0.7^{+1.9}_{-0.62}$   & 0  &   3.42 & 976 & 0.0035 \\\vspace{5pt}
    5   & $0.06^{+0.25}_{-0.06}$ & 0  &   2.59 & 1141 & 0.0023 \\\vspace{5pt}
    10  & $0.02^{+0.06}_{-0.02}$ & 0  &   2.62 & 1263 & 0.0021 \\ 
\bottomrule
\end{tabular}
\caption{Upper limits on the integral diffuse flux of UHE photons (last column). The different energy thresholds are listed in the first column. The following columns refer to 
    the expected number of background events $N_\text{b}$ along with its uncertainty $\sigma_\text{b}$, the number of photon candidates $N_\upgamma$, 
    the 95\% upper limits and the weighted exposure.}
\label{08::tab::ul::nBackground2}
\end{table*}

The limits derived from different analyses published by the Pierre Auger collaboration, as well as
those published by other cosmic ray observatories, are illustrated in \cref{fig:PhotonLimits}.
For comparison, in \cref{fig:PhotonLimits} the expected fluxes of ultra-high-energy photons based on different assumptions are also presented. 
Expectations for two distinct pure-proton scenarios \cite{Sarkar:2011hkm, universe8080402} are plotted, along with a scenario involving a mixed composition at the sources \cite{Bobrikova:2021wG}. 
While experimental sensitivities above \(3\times 10^{18}\)\,eV are approaching or already constraining the optimistic expectations for the photon flux produced by the interaction of protons during propagation, 
they remain approximately 1 to 1.5 orders of magnitude above those derived for the mixed-composition model.

Previous upper limits on the photon flux have constrained non-acceleration models, especially Super-Heavy Dark Matter (SHDM) models attempting to elucidate the origin of cosmic rays at the highest energies (see, e.g., \cite{PierreAugerPh:2017, *PierreAugerPh:2017_erratum, PierreAugerPh:2023}). 
The upper limits on the incoming photon flux allow for the constraint of the mass and lifetime phase space of SHDM particles~\cite{ANCHORDOQUI2021102614,PhysRevD.107.042002,PhysRevLett.130.061001}.
In \cref{fig:PhotonLimits}, expectations for three distinct assumptions regarding SHDM decay are displayed. 
For the hadronic decay (\(X \rightarrow q\bar{q}\)), expected fluxes are shown according to \cite{PhysRevD.94.063535} for SHDM particle masses (\(M_X\)) of \(10^{18}\) GeV with a lifetime (\(\tau_X\)) of \(3\times 10^{21}\)\,yr, 
as well as for \(M_X = 10^{12}\) GeV with \(\tau_X = 10^{23}\) yr, both combinations are currently within permissible bounds. 
Considering the decay into leptons, 
the expected flux according to \cite{PhysRevD.98.083016} is presented for \(M_X = 10^{10}\) GeV with \(\tau_X = 3 \times 10^{21}\) yr. 
With the increasing sensitivity of current photon searches, further constraints on these values become achievable.

\subsection{Systematic Uncertainties}
Various sources of systematic uncertainties in the calculated upper limits were investigated. 
The main contributions can be attributed to the systematic uncertainties on the energy scale~\cite{spectrumPRD} and on the \xmaxT reconstruction~\cite{PierreAuger:2014sui}.
In particular, shifting all energy values upward or downward by 14\%,
the number of candidates changes by $^{+5}_{-8}$ above the lowest energy threshold of 1 EeV, and by $^{+3}_{-1}$ above 2 EeV, while there is no change for the higher energy thresholds.
Similarly, shifting 
the reconstructed \xmaxT values 
by $\Delta\xmaxM=\pm10$\,g~cm$^{-2}$ the number of candidates changes by $^{+6}_{-2}$ above 1 EeV, while the higher energy intervals 
are not affected.
The discussed systematic shifts would have a similar impact on the expected background, as it is extrapolated from the burnt sample.
The influence of these systematic uncertainties 
on photon selection efficiency was also evaluated, leading to a contribution
of about $5\%$ from the uncertainties on the energy scale and $\sim 14\%$ from the uncertainties on the \xmaxT reconstruction.
The systematic uncertainty in the calculation of the hybrid exposure, estimated at the level of 6.4\% (see \cref{sec::exposure}), would propagate linearly into an additional systematic uncertainty on upper limits . 



Another source of uncertainty is the unknown photon spectral index, which would reflect into a change in the exposure. Differences of 15\% and 20\% are found in the first two energy intervals when changing the
spectral index from 2 to 1.5 and 2.5, respectively.
The lack of knowledge of $\Gamma$ may also have an impact on the analysis, because a different spectral index changes the shape of the distributions used as input variables for the MVA method. However, in the case of the Fisher Discriminant Analysis, the impact of the change in shape has been found to be negligible compared to the exposure effect.

The choice of a hadronic interaction model can significantly impact the differentiation between photons and protons, given that various models provide unique predictions for \xmaxT and the number of muons in showers initiated by hadronic primaries. 
The uncertainties inherent in modeling proton- and nucleon-induced air showers, which are pivotal for our analysis, may consequently influence the Fisher discriminant analysis.
In this study, we employed EPOS-LHC as the designated hadronic interaction model. Air showers generated using this model exhibit the highest substantial muon component and the deepest \xmaxT when compared to alternative models. This particular characteristic leads to more conservative upper limits, as illustrated in~\cite{PierreAugerPh:2017,*PierreAugerPh:2017_erratum}.
The effects of a different hadronic model have been investigated by considering Sybill\,2.3c simulations.
Notably, we observed a variation of the upper limits of $-14$\%, $-8$\%, $-6$\%, $-2$\%, and $+2$\% at 1, 2, 3, 5, and 10 EeV, respectively.

\section{Conclusions}
\label{sec::conclusions}
In this study, we performed a search for EeV photons using the full hybrid data sample collected by the Pierre Auger Observatory. The analysis combines the depth at the shower maximum, \(X_{\text{max}}\), directly measured from the Observatory's fluorescence detector,  with a parameter related to the muonic component of a shower, $F_{\mu}$, derived from signals of the surface detector, exploiting the air-shower universality paradigm.

The photon selection identified 22 candidates, consistent with the expected background and its uncertainty. 
This result, summarized in \cref{08::tab::ul::nBackground2}, establishes the most stringent upper limits on the diffuse UHE-photon flux above various energy thresholds. The limits, determined at a 95\% confidence level, are 0.0403, 0.01113, 0.0035, 0.0023, and 0.0021 km$^{-2}$ sr$^{-1}$ yr$^{-1}$ at 1, 2, 3, 5, and 10 EeV, respectively.


Compared to previous 
analyses~\cite{PierreAugerPh:2017, PierreAugerPh:2017_erratum}, 
the reported upper limits exhibit a substantial improvement, up to a factor of $\sim 3$ in the energy region above 3 EeV, where no candidates were found and the background is compatible with zero. 

This result can be attributed to an increase of about 50\% in measurement time in combination with a near doubling of the event selection efficiency, leading to a three-fold increase of the exposure.
Remarkably, the upper limit above 10 EeV turns out to be at the level of the limit obtained with the surface detector in the corresponding energy range~\cite{PierreAugerPh:2023}.   
Finally, the enhancement above the lowest energy threshold is about 40\%, primarily attributed to the inclusion of a background estimate in the calculation of the upper limits.

While the current limits do not challenge the flux of  photons produced during the propagation of UHECRs under the assumption of a mixed composition, they begin to probe the most optimistic predictions of pure-proton scenarios. 
Moreover the upper limits on the incoming photon flux allow for the constraint of the mass and lifetime phase space of super-heavy dark matter particles. 
Finally, the analysis presented in this study can serve as a complementary method for directional searches from specific targets or searches in  coincidence with observations from other cosmic messengers such as neutrinos and/or gravitational waves.
Future data will enhance the ability to constrain different mechanisms expected to produce UHE photons. 
Specifically, the completion of the upgrade of the Pierre Auger Observatory, AugerPrime~\cite{thepierreaugercollaboration2016, castelli2019}, is expected to further increase the sensitivity of the analysis.

\appendix
\section{Background estimation}
\label{sec::appendixA}
The goal of this section is to describe the distribution of the Fisher discriminant for the background. This is achieved in two steps as shown in~\cite{Savina:2021xpt}.
In the first step, we study its shape by profiting of the statistics offered by the proton simulations. Only the rightmost tail of the Fisher distribution is considered, specifically only the events with a Fisher discriminant \mbox{$\fisherM > -1.3$}, indicated by the blue vertical line in~\cref{fig:unblinding:a}.
This value of the Fisher discriminant is used because below $\fisherM_0 = -1.3$ the photon selection
efficiency is almost \SI{100}{\percent}.

\begin{figure}[ht]
  \centering
    \includegraphics[width=\columnwidth]{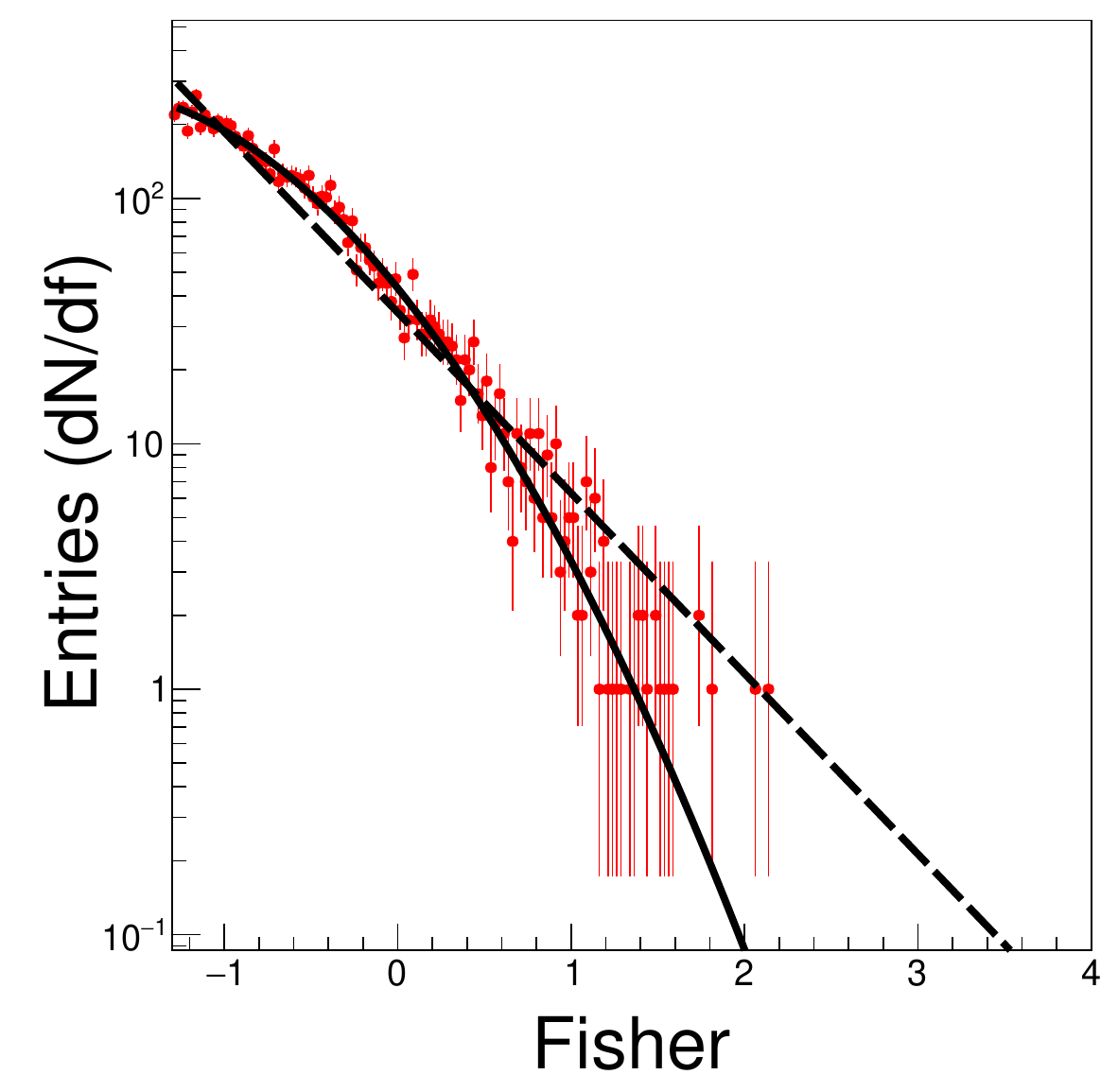}
    \caption{%
    Tail of the Fisher distribution for protons. 
    The two black lines represent the fits of the two functions,  $m^{\prime}$ (dashed) and $m$ (solid), 
    discussed in the text.}
    \label{07::fig::model::Fisher}
\end{figure}
\begin{figure}[ht]
    \centering
    \includegraphics[width=\columnwidth]{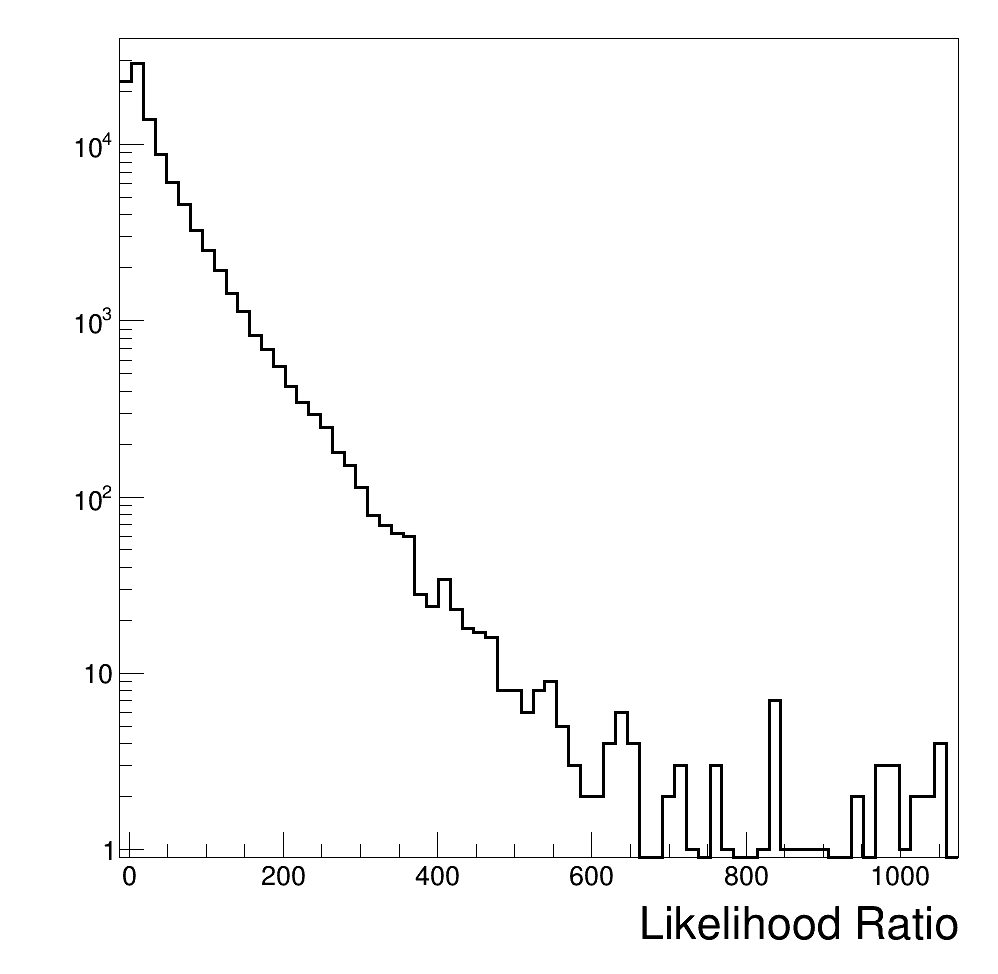}
  \caption{%
    Likelihood-ratio distribution from the fit of $m^{\prime}$ and $m$ to $10^6$ simulated samples (see text for details).}
    \label{07::fig::model::Likelihood}
\end{figure}

The tail of the proton distribution is highlighted in~\cref{07::fig::model::Fisher}, where two exponential functions, $m^{\prime}$ and $m$, tested for its description, are superimposed (black dashed and solid lines, respectively): 
\begin{subequations}
  \label{A::eq::model::background}
  \begin{align}
    m^{\prime}(\fisherM |B) &= N^{\prime}(B) e^{-B\fisherM} \label{A::eq::model::f1}\\
    m(\fisherM |A,B) &= N(A,B) e^{-(A\fisherM^{2}+B\fisherM)} \label{A::eq::model::f2}
  \end{align}
\end{subequations}
where $A$ and $B$ are shape parameters,
and $N^{\prime}$ and $N$ are the normalizations of $m^{\prime}$ and $m$, respectively. 
$N^{\prime}$ and $N$ are calculated as a function of the parameters $A$ and $B$,
by requiring that the integral of $m^{\prime}$ and $m$ is equal to the number of events $N_0$ that have a value of the Fisher discriminant above $\fisherM_0$, thus obtaining
\begin{subequations}
  \label{07::eq::model::normalization}
  \begin{align}
    N^{\prime}(A) &= \frac{N_0 B}{e^{-B\fisherM_0}} \label{07::eq::model::N1}\\
    N(A, B) &= \frac{N_0\sqrt{A}}{
                  e^{B^2/4A}\text{erfc}\left( \frac{B}{2\sqrt{A}} \left( \frac{2A}{B}\fisherM_0-1 \right)  \right)
                  }
                  \label{07::eq::model::N2}
  \end{align}
\end{subequations}
where erfc is the complementary error function.
The parameters obtained from an unbinned likelihood fit of $m^{\prime}$ and $m$ to the tail of the Fisher distribution are reported in~\cref{07::tab::model::parameters}.

\begin{table}[ht]
  \centering
  \begin{tabular}{ccc}
    \toprule
    &  $\mathbf{A}$  & $\mathbf{B}$ \\
    \midrule
    $m^{\prime}$  &   & 1.55 \\
    $m$  & 0.42 & -1.73\\
    \bottomrule
  \end{tabular}
  \caption{Values of the parameters $A$ and $B$ obtained from an unbinned likelihood fit to the tail of the Fisher distribution of protons, i.e., to events with $\fisherM>-1.3$.}
  \label{07::tab::model::parameters}
\end{table}

The best-fit model is determined by using a likelihood-ratio test~\cite{Lista:2016tva}, in which
two hypotheses on the shape of the tail distribution are compared: the null-hypothesis, $H_0$, according to which it
is described by $m^{\prime}$, i.e., $m^{\prime}(\fisherM |B) = m(\fisherM |A=0,B)$;
the alternative hypothesis, $H_{1}$, according to which it is described by $m(\fisherM |A \ne 0,B)$. The likelihood ratio $\mathcal{L}_{\text{ratio}}$ results to be $\approx 4000$. The p-value, $p_{\text{value}}$, associated to $\mathcal{L}_{\text{ratio}}$ is derived by applying the likelihood-ratio test on simulated samples of Fisher values, generated according to the $m^{\prime}$ model and then fitted with both models. Each sample consists of 30000 events (\textit{realizations}). The resulting distribution of the likelihood-ratios, based on \num{1000000} realizations, is shown in~\cref{07::fig::model::Likelihood}. As the maximum value attained in $10^6$ trials is about $1000$, i.e., $p_{\text{value}}(1000)\approx 10^{-6}$, then $p_{\text{value}}<10^{-6}$, i.e., the $m^{\prime}$ model is discarded in favor of $m$.

\begin{figure}[!ht]
    \centering
    \includegraphics[width=\columnwidth]{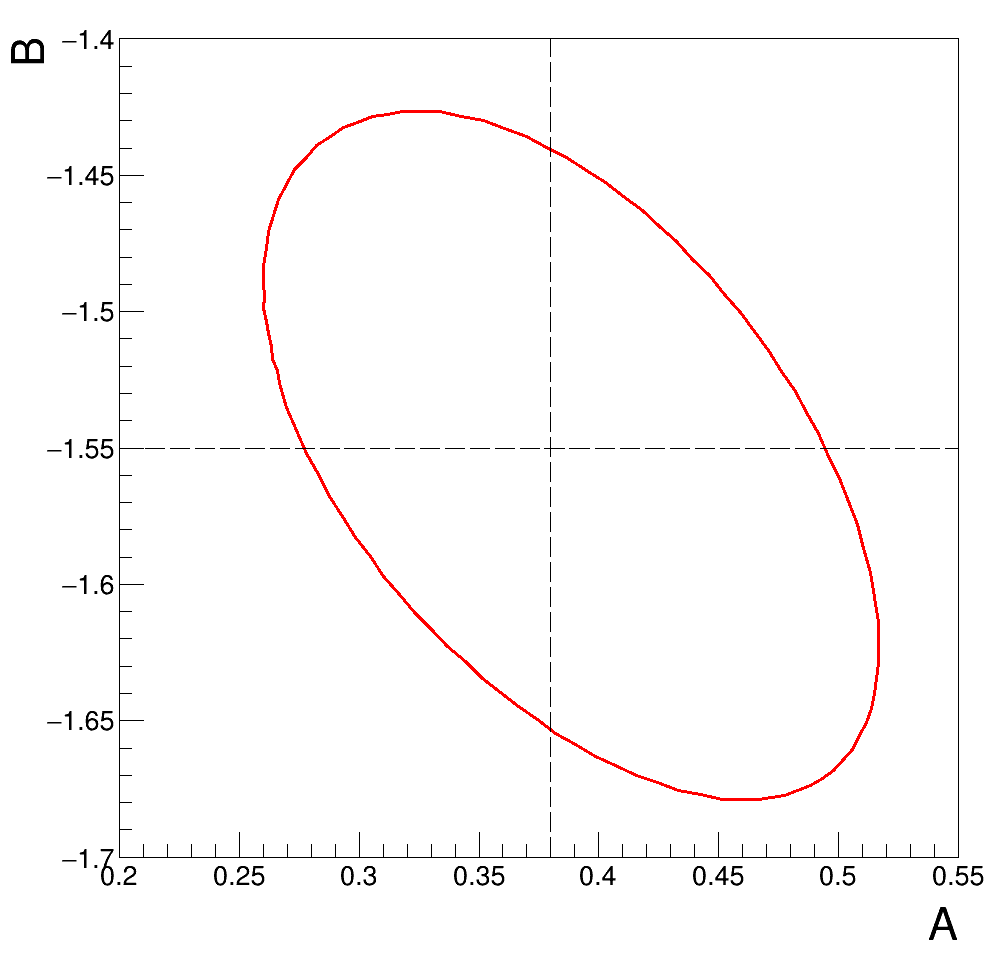}
  \caption{%
    1-sigma contour plot of the errors of the parameters $A$ and $B$.
    The dashed lines indicate the values obtained from the fit of $m$ on the burnt sample.
  }
  \label{07::fig::model::errors}
\end{figure}

\begin{figure}[ht]
  \centering
    \centering
    \includegraphics[width=\columnwidth]{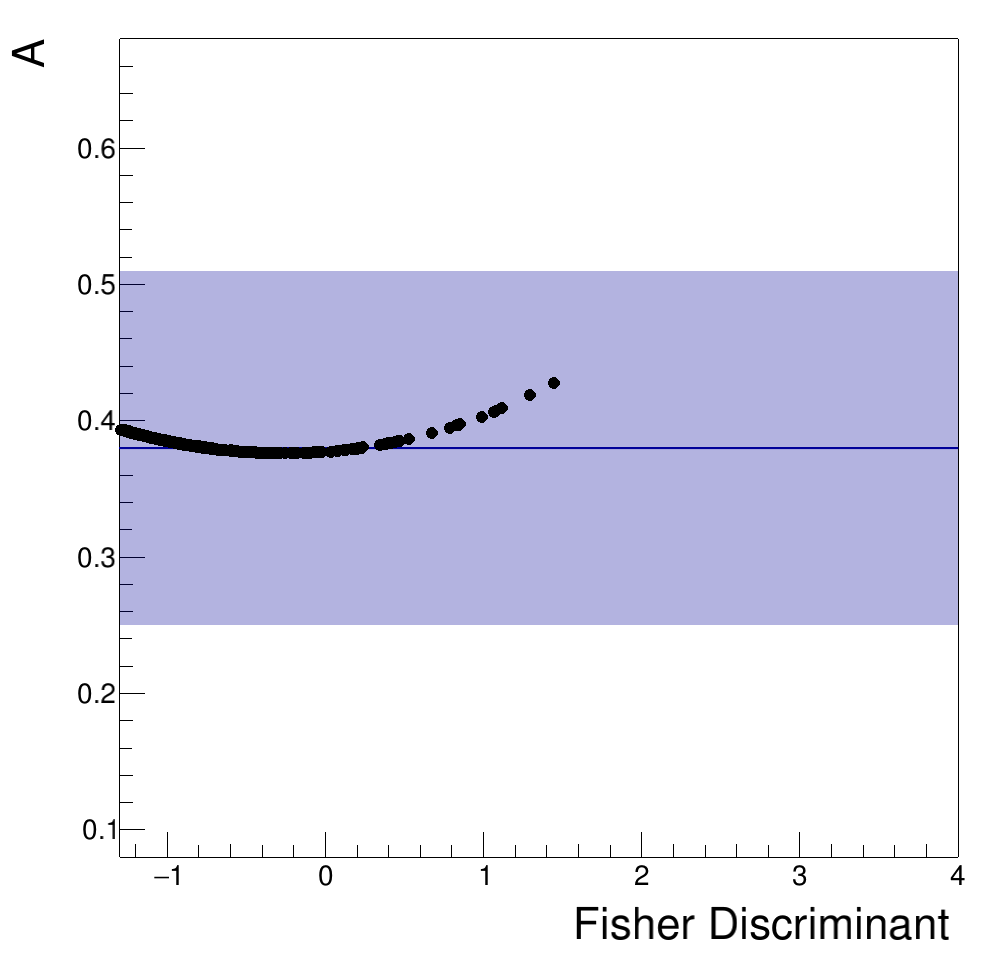}
    \caption{   Parameter $A$ as calculated with the jackknife technique (see text), as a function of the Fisher discriminant. The blue line shows the values obtained from the fit
    on the burnt sample, while the blue shaded area show the statistical uncertainties.}
    \label{07::fig::model::systA}
\end{figure}
\begin{figure}[ht]
    \centering
    \includegraphics[width=\columnwidth]{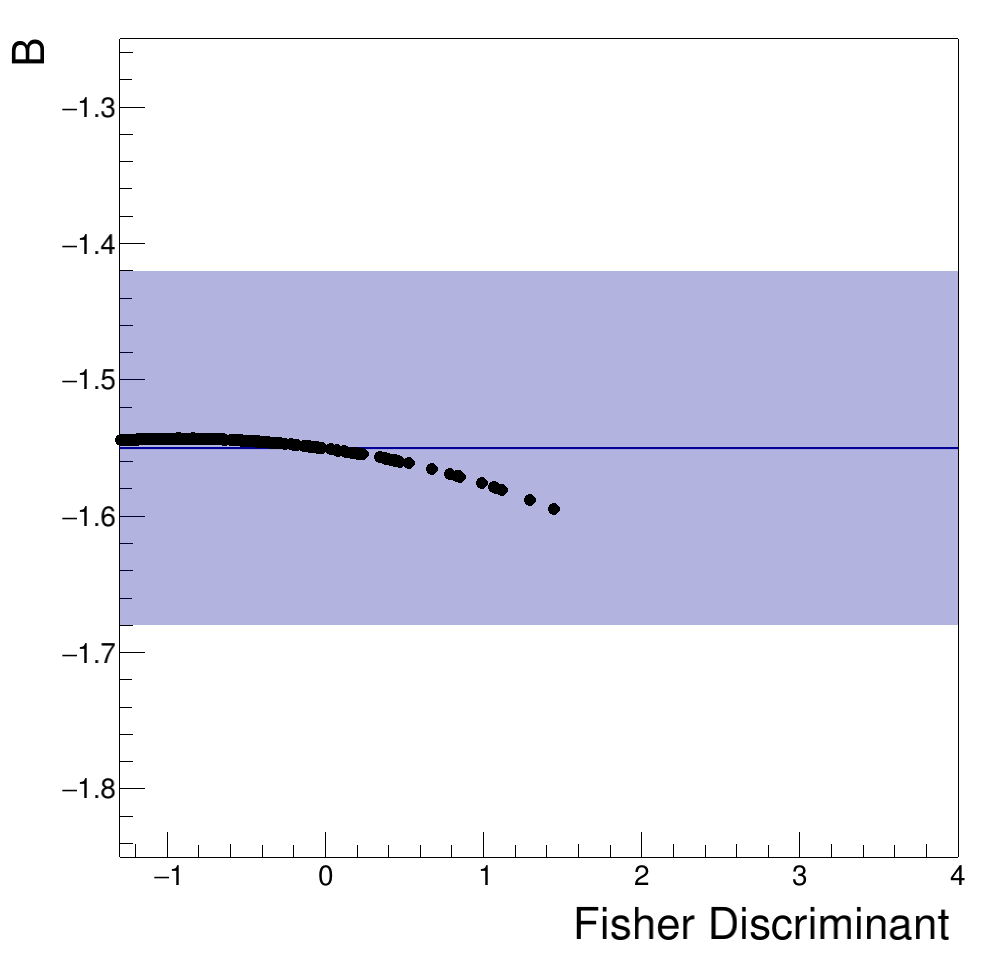}
    \caption{    Parameter $B$ as calculated with the jackknife technique (see text), as a function of the Fisher discriminant. The blue line shows the values obtained from the fit
    on the burnt sample, while the blue shaded area show the statistical uncertainties.}
    \label{07::fig::model::systB}
\end{figure}

After having derived the shape of the background from proton simulations, the second step in the characterisation of the background involves the burnt sample. 
To finalize the estimation of the background, a fit with the $m$ model is performed on the burnt sample distribution, as shown in~\cref{fig:background} (see \cref{sec::background}). 
The best-fit values of $A$ and $B$ are $0.38$ and $-1.55$, respectively: they are represented by the grey dashed lines in~\cref{07::fig::model::errors}, together with the red ellipses that marks the 1-sigma contour of the statistical errors. 

A possible photon contamination in the burnt sample cannot however be excluded: a possibly related systematic effect has thus been studied by using a jackknife technique~\cite{JKEfron}.
This is a re-sampling technique, which involves a leave-one-out strategy for the
estimation of the parameters (in this case, $A$ and $B$)
in a data set of $N$ observations.
The values of $A$ and $B$ calculated as a function of the Fisher discriminant \fisherT with this technique are shown in \cref{07::fig::model::systA} and \cref{07::fig::model::systB}, respectively. The blue shaded area represents the statistical
uncertainties obtained from the fit. 
As one can see, the systematic deviations from the central values (marked by the blue lines) are negligible with respect to the statistical ones. A systematic uncertainty of $0.01$ is then derived from the width of the distribution of the deviations when projected on the y-axis.

The parameters of the burnt sample obtained from the fit of the model $m$ to the burnt sample distribution are thus:
\begin{subequations}
  \label{A::eq::model::parameters}
  \begin{align}
    A = \ \  0&.38 \pm \text{(stat)} \ 0.13 \pm \text{(syst)} \ 0.01\\
    B = -1&.55 \pm \text{(stat)} \ 0.13 \pm \text{(syst)} \ 0.01
  \end{align}
\end{subequations}

Finally, to extrapolate the parametrisation of the Fisher distribution of the background to the full hybrid data set, the normalization of the function $m$ is scaled to the number of total events by setting $ N = N_{\text{data}}$ in~\cref{07::eq::model::N2}, where $N_{\text{data}} = 1328$.
The distribution of the Fisher discriminant for the extrapolated background is shown as a blue line in~\cref{fig:background}.
The uncertainties, $\sigma_{f}$, in the extrapolation, represented by the blue band, is calculated as
\begin{equation}
  \label{07::eq::model::sigmaf}
  \sigma_{f} = \sum_{i,j = A, B}\frac{\partial m}{\partial i} k_{ij}
  \frac{\partial m}{\partial j} 
\end{equation}
where $i$ and $j$ runs over the parameters $A$ and $B$, and $k_{ij}$ are the elements of the covariance
matrix:
\begin{equation}
  \label{07::eq::model::covarianceMatrix}
  K =
  \begin{pmatrix}
    0.0165 & -0.0086 \\
    -0.0086 & 0.0158 
  \end{pmatrix}
\end{equation}

\section{The most significant photon candidate event}
In this appendix the characteristics of the most significant photon candidate event (ID 3478968, see~\cref{sec::results}, and~\cref{08::tab::candidates::candidates}) are discussed in detail. The event has occurred on 22 May 2007, arriving at a zenith angle of about 57$^\circ$ and with reconstructed energy $E_\upgamma = (1.73 \pm 0.16){\times}10^{18}$\,eV, and 
$\xmaxM = (1245 \pm 57)$\,g~cm$^{-2}$.

The hybrid event, detected with both the surface and fluorescence detectors simultaneously, triggered 6 stations of the ground array, providing independent SD trigger and reconstruction. The FD telescope, Los Leones, observed the event that triggered 14 pixels, with a total angular track of around 17 degrees. The event passed all the selection criteria of the analysis and had the highest Fisher value in the data sample, 2.87.

\begin{figure*}[hbt]
  \centering
    \centering
    \includegraphics[width=0.95\textwidth]{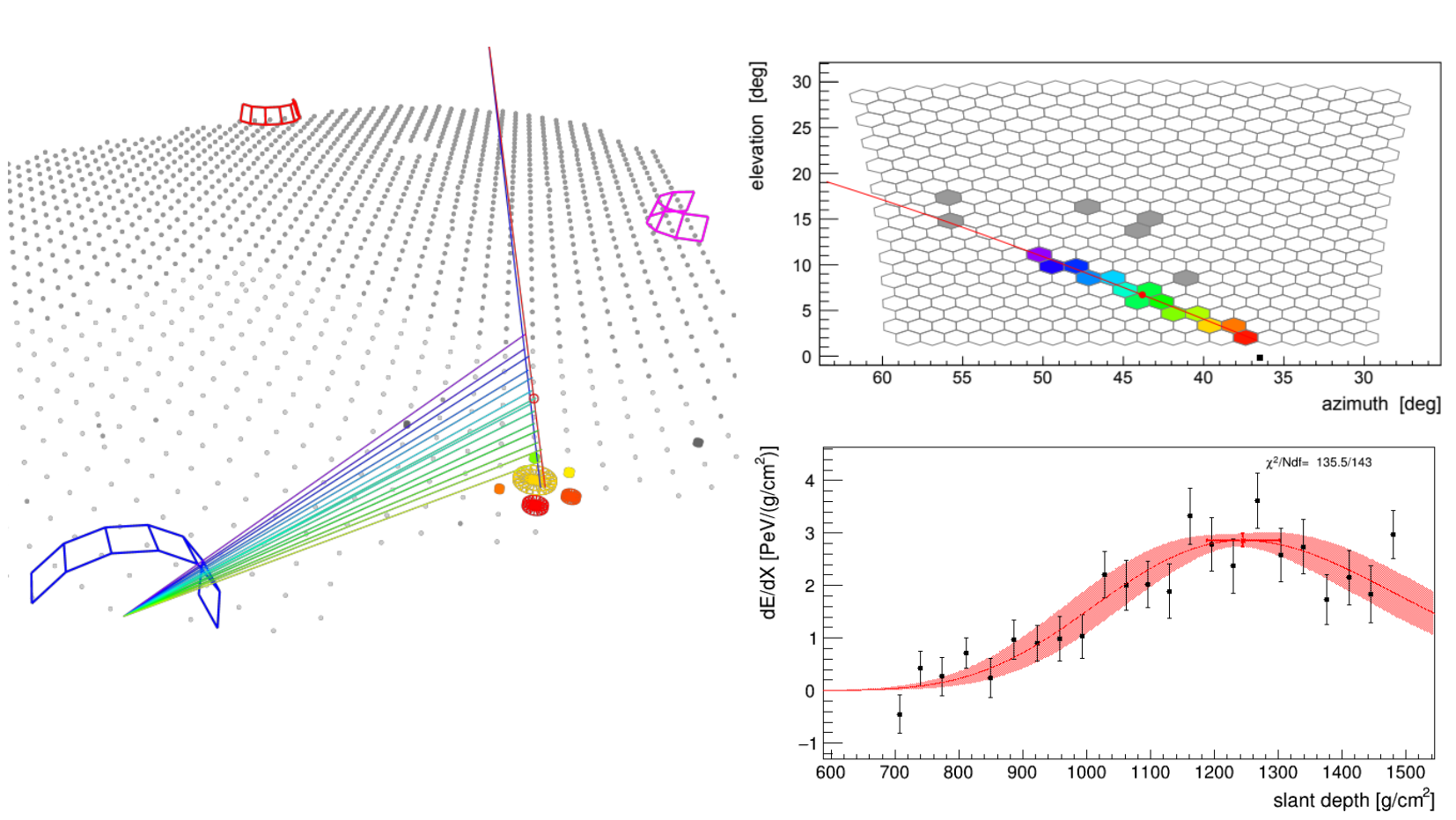}
   \caption{Characteristics of the most significant photon candidate event, ID 3478968: 3D visualization (left panel), 
    camera view of the triggered FD telescope (top right panel). The colors (violet to red) reflect the times (early to late) 
    at which the light reaches each pixel. Gray pixels indicate background triggered pixels. Bottom right panel, reconstructed energy deposit as a function of atmospheric slant depth (black points) along with the fit to the shower longitudinal profile (red line).
    }
    \label{AB:photon_candidate}
\end{figure*}

In~\cref{AB:photon_candidate}, left panel, the 3D visualization of the event at the Observatory is provided. The fluorescence telescopes are represented by colored squares, the ground array stations are marked with gray dots. The line of sight of the triggered FD pixels looking towards the reconstructed shower axis are shown as colored lines, and the circles show the positions of the triggered SD stations. Colors reflect triggering times: violet corresponds to early and red to late times. 
The camera view of the FD telescope detecting the event is shown in~\cref{AB:photon_candidate}, top right panel: the horizontal axis corresponds to the azimuth angle in the FD site local system while the vertical axis is the angular elevation of the viewing direction of the FD pixels. Same color code for triggering times, gray pixels are background triggered pixels not participating in the event geometry reconstruction.
The energy deposit as a function of the traversed atmospheric slant depth is shown in~\cref{AB:photon_candidate}, bottom right panel. The fit to the shower longitudinal profile (red solid line) and its uncertainty (red band) are shown in the Figure along with the position of the shower maximum and its uncertainty (red point).

\section*{Acknowledgments}

\begin{sloppypar}
The successful installation, commissioning, and operation of the Pierre
Auger Observatory would not have been possible without the strong
commitment and effort from the technical and administrative staff in
Malarg\"ue. We are very grateful to the following agencies and
organizations for financial support:
\end{sloppypar}

\begin{sloppypar}
Argentina -- Comisi\'on Nacional de Energ\'\i{}a At\'omica; Agencia Nacional de
Promoci\'on Cient\'\i{}fica y Tecnol\'ogica (ANPCyT); Consejo Nacional de
Investigaciones Cient\'\i{}ficas y T\'ecnicas (CONICET); Gobierno de la
Provincia de Mendoza; Municipalidad de Malarg\"ue; NDM Holdings and Valle
Las Le\~nas; in gratitude for their continuing cooperation over land
access; Australia -- the Australian Research Council; Belgium -- Fonds
de la Recherche Scientifique (FNRS); Research Foundation Flanders (FWO),
Marie Curie Action of the European Union Grant No.~101107047; Brazil --
Conselho Nacional de Desenvolvimento Cient\'\i{}fico e Tecnol\'ogico (CNPq);
Financiadora de Estudos e Projetos (FINEP); Funda\c{c}\~ao de Amparo \`a
Pesquisa do Estado de Rio de Janeiro (FAPERJ); S\~ao Paulo Research
Foundation (FAPESP) Grants No.~2019/10151-2, No.~2010/07359-6 and
No.~1999/05404-3; Minist\'erio da Ci\^encia, Tecnologia, Inova\c{c}\~oes e
Comunica\c{c}\~oes (MCTIC); Czech Republic -- GACR 24-13049S, CAS LQ100102401,
MEYS LM2023032, CZ.02.1.01/0.0/0.0/16{\textunderscore}013/0001402,
CZ.02.1.01/0.0/0.0/18{\textunderscore}046/0016010 and
CZ.02.1.01/0.0/0.0/17{\textunderscore}049/0008422 and CZ.02.01.01/00/22{\textunderscore}008/0004632;
France -- Centre de Calcul IN2P3/CNRS; Centre National de la Recherche
Scientifique (CNRS); Conseil R\'egional Ile-de-France; D\'epartement
Physique Nucl\'eaire et Corpusculaire (PNC-IN2P3/CNRS); D\'epartement
Sciences de l'Univers (SDU-INSU/CNRS); Institut Lagrange de Paris (ILP)
Grant No.~LABEX ANR-10-LABX-63 within the Investissements d'Avenir
Programme Grant No.~ANR-11-IDEX-0004-02; Germany -- Bundesministerium
f\"ur Bildung und Forschung (BMBF); Deutsche Forschungsgemeinschaft (DFG);
Finanzministerium Baden-W\"urttemberg; Helmholtz Alliance for
Astroparticle Physics (HAP); Helmholtz-Gemeinschaft Deutscher
Forschungszentren (HGF); Ministerium f\"ur Kultur und Wissenschaft des
Landes Nordrhein-Westfalen; Ministerium f\"ur Wissenschaft, Forschung und
Kunst des Landes Baden-W\"urttemberg; Italy -- Istituto Nazionale di
Fisica Nucleare (INFN); Istituto Nazionale di Astrofisica (INAF);
Ministero dell'Universit\`a e della Ricerca (MUR); CETEMPS Center of
Excellence; Ministero degli Affari Esteri (MAE), ICSC Centro Nazionale
di Ricerca in High Performance Computing, Big Data and Quantum
Computing, funded by European Union NextGenerationEU, reference code
CN{\textunderscore}00000013; M\'exico -- Consejo Nacional de Ciencia y Tecnolog\'\i{}a
(CONACYT) No.~167733; Universidad Nacional Aut\'onoma de M\'exico (UNAM);
PAPIIT DGAPA-UNAM; The Netherlands -- Ministry of Education, Culture and
Science; Netherlands Organisation for Scientific Research (NWO); Dutch
national e-infrastructure with the support of SURF Cooperative; Poland
-- Ministry of Education and Science, grants No.~DIR/WK/2018/11 and
2022/WK/12; National Science Centre, grants No.~2016/22/M/ST9/00198,
2016/23/B/ST9/01635, 2020/39/B/ST9/01398, and 2022/45/B/ST9/02163;
Portugal -- Portuguese national funds and FEDER funds within Programa
Operacional Factores de Competitividade through Funda\c{c}\~ao para a Ci\^encia
e a Tecnologia (COMPETE); Romania -- Ministry of Research, Innovation
and Digitization, CNCS-UEFISCDI, contract no.~30N/2023 under Romanian
National Core Program LAPLAS VII, grant no.~PN 23 21 01 02 and project
number PN-III-P1-1.1-TE-2021-0924/TE57/2022, within PNCDI III; Slovenia
-- Slovenian Research Agency, grants P1-0031, P1-0385, I0-0033, N1-0111;
Spain -- Ministerio de Ciencia e Innovaci\'on/Agencia Estatal de
Investigaci\'on (PID2019-105544GB-I00, PID2022-140510NB-I00 and
RYC2019-027017-I), Xunta de Galicia (CIGUS Network of Research Centers,
Consolidaci\'on 2021 GRC GI-2033, ED431C-2021/22 and ED431F-2022/15),
Junta de Andaluc\'\i{}a (SOMM17/6104/UGR and P18-FR-4314), and the European
Union (Marie Sklodowska-Curie 101065027 and ERDF); USA -- Department of
Energy, Contracts No.~DE-AC02-07CH11359, No.~DE-FR02-04ER41300,
No.~DE-FG02-99ER41107 and No.~DE-SC0011689; National Science Foundation,
Grant No.~0450696; The Grainger Foundation; Marie Curie-IRSES/EPLANET;
European Particle Physics Latin American Network; and UNESCO.
\end{sloppypar}

\bibliography{main}

\end{document}